\documentclass[12pt]{article}
\usepackage{amsmath}
\usepackage{euscript}
\usepackage[T1]{fontenc}
\usepackage[latin1]{inputenc}
\usepackage{graphics}
\usepackage{longtable}
\usepackage{color}
\usepackage{pslatex}
\usepackage{subfigure}
\usepackage{hyperref}

\usepackage{pstricks,epsfig,amssymb}
\usepackage{a4wide, amsmath, latexsym, epsfig,
  psfrag,graphicx, rotating, fancyhdr, tabularx, afterpage}

\usepackage{cite}  
\begin{document}

\def\lb{\nextline}
 
\def\Order#1{{\cal O}($#1$)}
\def\Kuhn{K\"uhn}
\def\alphapi{\Bigl({\alpha\over\pi}\Bigr)}
\def\sovem{{s\over m^2_e}}
\def\Born{{\rm Born}}
\def\nubar{\bar{\nu}}
\def\nubarnu{\bar{\nu}\nu}
\def\nubart{\bar{\nu}_\tau}
\def\sstrut{$\strut\atop\strut$}

  \def\PL #1 #2 #3 {Phys. Lett. {\bf#1}           (#3)  #2}
  \def\NP #1 #2 #3 {Nucl. Phys. {\bf#1}           (#3)  #2}
  \def\PR #1 #2 #3 {Phys. Rev. {\bf#1}            (#3)  #2}
  \def\PP #1 #2 #3 {Phys. Rep. {\bf#1}            (#3)  #2}
  \def\PRL #1 #2 #3 {Phys. Rev. Lett. {\bf#1}     (#3)  #2}
  \def\CPC #1 #2 #3 {Comp. Phys. Commun. {\bf#1}  (#3)  #2}
  \def\ANN #1 #2 #3 {Annals of Phys. {\bf#1}      (#3)  #2}
  \def\APP #1 #2 #3 {Acta Phys. Pol. {\bf#1}      (#3)  #2}
  \def\ZP  #1 #2 #3 {Z. Phys. {\bf#1}             (#3)  #2}

\def\uncatcodespecials{\def\do##1{\catcode`##1=12 }\dospecials}
\def\setupverbatim{\tt
  \def\par{\leavevmode\endgraf} \catcode`\`=\active
  \obeylines \uncatcodespecials \obeyspaces \parindent=5mm \parskip=0pt}
{\obeyspaces\global\let =\ } 
{\catcode`\`=\active \gdef`{\relax\lq}}
\def\beginverbatim{\par\begingroup\setupverbatim\doverbatim}
{\catcode`\|=0 \catcode`\\=12 
  |obeylines|gdef|doverbatim^^M#1\endverbatim{#1|endgroup}}
\def\Was{\hbox{W\c as\;}}
\def\M{\hbox{\cal M}}
\def\lips{\hbox{Lips}}
\def\Im{\hbox{Im}}
\def\GeV{\hbox{GeV}}
\def\Maj{M_{R}}  \def\Gaj{\Gamma_{R}}
\def\beq{\begin{equation}} 
\def\eqiv{\sim}
\def\eeq{\end{equation}} 
\def\eps{\epsilon}
\begin{titlepage}
 
\begin{flushright} 
{ IFJPAN-IV-2009-1 \\ CERN-PH-TH/2009-092} 
\end{flushright}
 
\vspace{0.2cm}
\begin{center}
{\bf\Large
Matching NLO parton shower matrix element with exact phase space:
case of $W\to l \nu $($\gamma$) and $\gamma^* \to \pi^+\pi^-$($\gamma$)$^{\dag}$
}\end{center}
 
\begin{center}
   {\bf  G. Nanava$^a$,   Qingjun Xu$^{b,c}$ and Z. W\c{a}s$^{c,d}$}\\
   {\em $^a$ Physikalisches Institut, Universit\"at Bonn,
       Nussallee 12, 53115 Bonn, Germany
        \\ (On leave from IHEP, TSU, Tbilisi, Georgia)
}\\ 
{\em $^b$ Department of Physics, Hangzhou Normal University, 
Hangzhou 310036, China  }\\
       {\em $^c$  Institute of Nuclear Physics, PAN,
        Krak\'ow, ul. Radzikowskiego 152, Poland}\\
{\em $^d$ CERN PH-TH, CH-1211 Geneva 23, Switzerland }
\end{center}
\vspace{.1 cm}
\begin{center}
{\bf   ABSTRACT  }
\end{center}

The 
{\tt PHOTOS} Monte Carlo is often used for simulation of QED effects in decay 
of intermediate particles and resonances. Momenta are 
generated in such a way that samples of events
cover the whole bremsstrahlung  phase space.   With the help of selection cuts,
experimental acceptance can  be then taken into account. 

The program is based on an exact 
multiphoton phase space.  Crude matrix element is obtained by iteration
of a universal multidimensional kernel. It ensures exact distribution 
in the soft photon region. Algorithm is compatible with exclusive exponentiation.
To evaluate the program's precision, it is necessary
to control the kernel with the help of  perturbative results. 
If availabe, kernel is constructed from the exact first order matrix element.
This ensures that  all terms necessary for non-leading 
logarithms are taken into account.
 In the present paper we will focus 
on the $W \to l \nu$ and $\gamma^* \to \pi^+ \pi^-$ decays. 
 The Born level cross 
sections for both processes approach zero in some points of the phase space.

A process dependent compensating weight is  
constructed
to incorporate the  exact matrix element, but is recommended for use in tests only.
In the hard photon region, where scalar QED is not expected to be reliable,
the compensating weight for $\gamma^*$ decay can be large.
With respect to the total rate, the effect 
remains at  the permille level. It is nonetheless of 
interest. The terms leading to the effect  are analogous to some terms appearing
in  QCD.

The present paper can be understood either as a contribution to 
discussion on how to 
match  two collinear emission  chains resulting  from charged sources 
in a way compatible with the exact and complete phase space, exclusive exponentiation
  and the first order
matrix element of QED (scalar QED),
or as the practical study of predictions for  accelerator experiments.

 
 \vspace{0.5cm}
\begin{flushleft}
{   IFJPAN-IV-2009-1 \\  CERN-PH-TH/2009-092 \\
 June, 2009}
\end{flushleft}
 
\vspace*{1mm}
\bigskip
\footnoterule
\noindent
{\footnotesize \noindent  $^{\dag}$
This work is partially supported by  EU Marie Curie Research Training Network 
grant under the contract No. MRTN-CT-2006-0355505,   
Polish Government grant N202 06434 (2008-2010) and 
 EU-RTN Programme: Contract
No. MRTN-CT-2006-035482, 'Flavianet'.}
\end{titlepage}

\section{ Introduction}

One of the crucial goals of any high energy physics experiments is the comparison
between results of new measurements and predictions obtained from theory.
If agreement is obtained, then the validity domain for the theory is extended.
Discrepancy can be attributed to the  so called new physics. 
This scheme is in principle rather 
simple, but in practice, it is 
involved. For LEP experiments, enormous effort for such a program was documented 
in \cite{Altarelli:367653,Jadach:2000zp}. It was necessary
because for  very precise scattering experiments one needs to study radiative 
corrections simultaneously with detector acceptance. As a consequence, it was
possible to confirm experimentally 
 that the Standard Model was indeed a field theory of elementary particle
interactions. Quantum effects could not be omitted, they had to be included 
in calculations. The
importance of this achievement was confirmed by the 1999 year Nobel Prize 
attributed to 't Hooft and Veltman. Also the 2008 Nobel Prize for the 
mechanism of 
quark flavour mixing \cite{NobelPrize2008} required precise measurements 
and comparison with  data.
These experiments, Belle and BaBar, located  respectively in Japan and USA 
required good control of radiative corrections too 
\cite{Abe:2006xs,Aubert:2002hc}. 

Because of the nature of accelerator experiments,
 it is generally believed that the Monte Carlo technique is the only suitable 
tool for the precision comparison 
of theory with experimental data \cite{Altarelli:367653}: 
effects of detector acceptance can be merged into theoretical predictions 
by simple rejection of some of the generated events. Theoretical
effects of different nature  can be taken into account, in particular 
radiative corrections. A multitude of Monte Carlo programs
were developed in context of QED \cite{Berends:1983mi,Jadach:1991ws} 
and QCD \cite{Sjostrand:1993yb,Corcella:2000bw}. The references can serve as 
examples.

Such Monte Carlo programs must rely on results obtained from 
 perturbative methods. In the 
case of QED, 
exponentiation  \cite{Yennie:1961ad}
is  useful. Exponentiation is a long established and 
rigorous scheme of reorganization 
of perturbative expansions. 
It was found \cite{Jadach:1987ii} that
 Monte Carlo programs can be developed using it as a basis. 
Significant theoretical effort was nonetheless necessary.
It required not only explicit calculation of exact fixed order cross sections,
but also to 
separate them into appropriate parts, at the  cross section or 
the spin amplitude level,
to finally  match results of fixed order calculations with
 coherent exclusive exponentiation (CEEX) \cite{Jadach:2000ir}
and  implement  it  into computer programs. 

QED predicts distributions which are  strongly peaked in  phase space.
They may vary by more than 10 orders of magnitude. This and the  complex 
structure of infrared singularity cancellations,  poses a challenge in Monte Carlo method. Appropriate
choice  of the crude
distribution over the phase space must be found. 
In the case of multi-photon radiation, a particularly elegant method was found
\cite{Jadach:2000ir}.
Thanks to conformal symmetry, it was possible to construct a crude distribution 
which was actually exact from the phase space point of view. All 
simplifications 
were localized in an approximated matrix element. An approximated matrix 
element consisting of the Born amplitude multiplied by the so-called
soft factors was  used at a first step.  Any further improvements could then 
be easily 
achieved with a  correcting weight. The weight
is the 
ratio of distributions calculated from an available
matrix element obtained perturbatively to a given fixed 
order and the one used in first step of the generation.

The case of QCD is by far more complex, but the general principle is similar.
One  constructs a simplified matrix element 
(and approximated  phase space) as a  basis  
 for the  parton shower algorithms. 
 Such a solution is limited to leading logarithms \cite{Kuraev:1976ge, Kuraev:1977fs,Balitsky:1978ic,Fadin:1998py,Ermolaev:2007yb}. 
Phase space organizations based on so-called orderings are often used 
\cite{Gorshkov:1966hu,Gorshkov:1966ht}.
Improvements beyond leading approximations 
are possible and widely used  \cite{Gribov:1984tu,Webber:1986mc},
but have
technical difficulties, for example
through the appearance of negative weight events. 

Another difficulty of QCD is the necessity to use
parameterization of $\alpha_s$ at low $Q^2$ where the standard
parameterization $\alpha_s = \alpha_s(Q^2)$ becomes 
unreliable\footnote{
Discussion of  this problem can be found in 
\cite{Ermolaev:2001zs, Ermolaev:2008dq}.}.
At  small scales non-perturbative aspects of QCD can dominate. 
Phenomena like
underlying event \cite{Affolder:2001xt} or hadronization
\cite{Derrick:1984mp} lead to further complications.
This is of course  on top
of hadron structure functions obtained from experiments
\cite{Berny:1989rf}.

We will not elaborate in our paper on these topics,  we think however, that 
the methods applied in this paper 
may provide a useful hint. 
First results \cite{vanHameren:2008dy} are encouraging.
Our paper is devoted  to   reliability 
of the {\tt PHOTOS} Monte Carlo.
{\tt PHOTOS} is a Monte Carlo \cite{Barberio:1990ms,Barberio:1993qi} 
for the QED bremsstrahlung in decays.
Its structure  is 
similar to the algorithms for QED exclusive exponentiation; 
the parameterization of its phase 
space  is exact, but its algorithm is iterative\footnote{That is why it is
 similar to solutions used in QCD parton showers, but no phase space
ordering  of any sort is applied and of course most of the difficulties
present in QCD are absent as well. One should also stress differences;
iterated, single emission kernel simultaneously feature all emission sources.
}. Conformal
symmetry is not used. This is advantageous, terms responsable
 for leading logarithms of  decay product are reproduced to all orders.  

As in the cases mentioned before
 effort in understanding results of exact perturbative
calculations was necessary for the  construction of {\tt PHOTOS}.  
Cross section level distributions, 
Refs. \cite{Berends:1982ie,RichterWas:1988gi}, were used. 
Later, thanks to experience gained in KKMC project \cite{Jadach:2000ir,Jadach:1999vf}, 
spin amplitudes were found to be helpful. In particular, results of
Refs. \cite{Was:2004ig,RichterWas:1994ep,RichterWas:1993ta} were 
used. They were essential for 
design and tests of the program, in particular for the choice of
single emission kernels. Thanks to these works, interference of consecutive emissions from a charged
line as well as interference of emissions from distinct charged lines
was  properly taken into account, without any need to divide the 
 phase space into differently treated sectors. That is also why
there was no need to separate the photon emission 
phase space into  regions where either a shower or a fixed order 
hard matrix element is used. 

Refs.
\cite{Golonka:2005pn,Golonka:2006tw,Nanava:2006vv}  were devoted 
 to numerical tests, but also a better explanation of
 theoretical foundation of  the program was  given there. 
It  may be worth mentioning that for many years the program's precision  was 
of no interest and such explanations were delayed  to  
the present decade.
 
The best detailed description of the phase space parameterization  
 as used in {\tt PHOTOS} and the explanation  that it is actually exact,
is given in Refs. \cite{Golonka:2006tw,Nanava:2006vv}. However,
it is not different from what was already explained in  
\cite{Barberio:1990ms,Barberio:1993qi}.

The precision of the program is significantly improved with respect to its
early versions. 
As it was shown in \cite{Golonka:2005pn,Golonka:2006tw}, even 
if the incomplete first order matrix element is used 
 in {\tt PHOTOS}, its results agree much  better with  {\tt KKMC}  using second order matrix
element exclusive exponentiation 
 \cite{Jadach:1999vf} than with {\tt KKMC},
using matrix element  restricted 
to first order and exponentiation. 
To quantify this statement, the
 method described in \cite{Golonka:2002rz, Davidson:2008ma} was used.
{\tt PHOTOS} was found to exploit result of
perturbative  calculation quite well, but it  cannot be a substitute
of such calculations.

The main goal of the present paper
is  to study spin amplitudes 
for construction of the process dependent 
weight, as in \cite{Golonka:2006tw,Nanava:2006vv},  but for
 $\gamma^* \to \pi^+\pi^-$ and  $W^\pm \to l^\pm \nu $. 
For that purpose, the  matrix element needs to be studied 
in great detail. Its gauge invariant parts\footnote{ In our case,
 gauge invariance reduces to independence of longitudinal component of photon polarization. }
need to be identified and with their help relations with amplitudes
of lower orders have to be found. This second aspect is  important
and is closely related to properties used in defining factorization schemes,
see eg. \cite{Gribov:1972ri,Collins:1985ue}. 

 Let us point out that in this paper we will not discuss spin amplitudes
from the perspective 
of matching consecutive emissions from the same charged line.
Such studies were performed earlier \cite{Barberio:1993qi} and for other decays;
these studies required double emission QED 
 amplitudes 
 \cite{RichterWas:1993ta,RichterWas:1994ep}. We will focus on single photon emission
and matching the  emissions from two charged lines in  
$\gamma^* \to \pi^+\pi^-$.
The analysis of the spin amplitudes and tests for the algorithm in the case of 
$Z$ decay into pair of
charged fermion was  given earlier, in ref. \cite{Golonka:2006tw}. 
The 
scalar particle decay into a pair of fermions was covered in \cite{Andonov:2002mx} and the decay of a 
spinless particle into a  pair of scalars was studied in \cite{Nanava:2006vv}. 
It seems that  the algorithm
works better (correction weights are less important) when 
initial state is spinless\footnote{This is particularly interesting from the
point of view of future attempts to extent into QCD. }. The case of $W$ decay was covered in \cite{Nanava:2003cg},
though some approximations were used and the decay requires to be revisited.

The two  processes are not only of the technical interest, they provide
examples 
for  studies of Lorentz and gauge group properties 
of spin amplitudes and cross sections. The
$\gamma^* \to \pi^+\pi^-$ decay is well measured.
It is important to improve theoretical
uncertainty of {\tt PHOTOS} for this decay, because of its relevance to
establishing $\alpha_{QED}(M_Z)$ and to
phenomenology of $g-2$.  From that perspective, the validity
 of our study is limited by validity of
scalar QED\footnote{ This last
constraint is of course common with the projects such as {\tt PHOKARA}
\cite{Rodrigo:2001kf}. {\tt PHOTOS} will not be better or worse from that point of view.}.
The $W \to l \nu$ decay is of interest for precision measurement of $W$ mass
and width, at LHC for example.

Our paper is organized as follows. In Section 2 we present the scalar QED
spin amplitudes for the process $e^+e^- \to \gamma^* \to \pi^+\pi^- (\gamma)$. 
It will be shown that the spin amplitudes can be separated
 into two gauge invariant parts.
Section 3 covers further discussion of the amplitudes, and the 
formulae for the cross section
used to obtain numerical results. A separation into eikonal part
and remaining parts is also presented.
Section 4 is devoted to the numerical results obtained with the help 
of {\tt MC-TESTER} \cite{Golonka:2002rz,Davidson:2008ma}. Different options of separating  non-leading effects are  
demonstrated.
Section 5 is devoted to the discussion of further tests, where 
distributions sensitive to the beam direction will
be used as well. Similarities and differences with respect to the
previous case will be underlined. 
Section 6 summarize the paper. Spin amplitudes for
$W^\pm \to l^\pm \nu $ are given in the Appendix.

\section{Amplitudes}
One of the necessary steps in the development of any Monte Carlo program is
 analyzing spin amplitudes calculated from  the theory or phenomenological model
under consideration. Fixed order analytical results are often not sufficient.
Even well known  amplitudes have to be revisited again to study
their structure.  It appears to be fruitful to study decompositions of the amplitude into sums
 of gauge invariant parts, which  can be further factorized into 
 gauge invariant terms.   In particular, to find correspondence with 
 factorization properties of the underlying field theory\footnote{The 
particulary rich case of $e^+e^- \to \nu_e \bar \nu_e \gamma$\;  
and  Monte Carlo implementation of spin amplitudes separated into parts  
is discussed in \cite{Bardin:2001vt,Was:2004ig}.}.

The spin amplitudes for $W\to l \nu \gamma$\; are collected in the Appendix. 
In principle they are straightforward and available already in appropriate form
in  \cite{Nanava:2003cg} but let us recall them again to clarify
 possible ambiguities on  how
emission from $W$ is separated into  final state radiation and initial
state radiation. Let us point also to  \cite{Berends:1984xv}, where spin amplitudes 
for radiative corrections and $W$ decay Monte Carlo are discussed for 
the first time.  
 This is an important starting point for the discussion of radiative 
corrections necessary for precise measurement of $W$ lineshape.

In the following, let us concentrate on the process
$\gamma^* (p) \rightarrow \pi^+(q_1) \pi^-(q_2) \gamma (k, \epsilon)$.
Since the precision required by experiments is lower in this case than at LEP, 
 we will limit
ourselves to the discussion  of amplitudes for single photon emission.
 We will not perform detailed analysis
of virtual corrections. At required level of precision, it is enough to anticipate their size 
thanks to the Kinoshita-Lee-Nauenberg theorem \cite{Kinoshita:1962ur,Lee:1964is}. 
Anyway, scalar QED predictions for our process are only partly reliable.

The spin structure of our process is new with respect to the processes we
have already 
studied: $Z \to l^+ l^- $, $h \to l^+ l^- $   
or $B^0 \to \pi^+ \pi^-$. The spin of the initial state can not be transmitted
into helicities of the outgoing particles. That is why we can expect different
properties of the amplitudes.
    
If one considers the process
$e^+e^- \to \gamma^* (p) \rightarrow \pi^+(q_1) \pi^-(q_2) \gamma (k, \epsilon)$,
amplitudes equivalent to those given in \cite{Czyz:2002np} are obtained.
The amplitude can be written as 
$M=V^\mu H_\mu$ where  $V_{\mu} =\bar{v}(p_1,\lambda_1) \gamma_\mu u(p_2,\lambda_2)$. The
$p_1,\lambda_1,p_2,\lambda_2$ are momenta and helicities of the incoming
electron and positron. The
 $V_{\mu}$ define the spin state of the intermediate $\gamma^*$.

Let us turn  to the  virtual photon decay now. Following conventions
of 
\cite{Czyz:2002np}, the final interaction part of the
Born matrix element for such process is 
\begin{equation}
H_0^\mu(p,q_1,q_2) = \frac{eF_{2\pi}(p^2)}{p^2}(q_1-q_2)^{\mu}.
\end{equation}
Here $p=q_1+q_2$. If a photon is present, this part 
of the amplitude reads:
\begin{equation}\label{eq:hmu1}
H^\mu =\frac{e^2F_{2\pi}(p^2)}{p^2}\left \{(q_1+k-q_2)^{\mu}
\frac{q_1\cdot \epsilon^*}{q_1\cdot k} +(q_2+k-q_1)^{\mu}
\frac{q_2\cdot \epsilon^*}{q_2\cdot k}-2\epsilon^{*\mu} \right\},
\end{equation}
which can be re-written to the following form:
\begin{eqnarray}\label{eq:bremsAmp0}
H^\mu 
& = &H_0^\mu(p,q_1,q_2) e \left (\frac{q_1\cdot \epsilon^*}{q_1\cdot k} - 
\frac{q_2\cdot \epsilon^*}{q_2\cdot k}\right ) +\nonumber \\
&&\frac{e^2F_{2\pi}(p^2)}{p^2}\left (\frac{k^{\mu}q_1\cdot \epsilon^* -
\epsilon^{*\mu}q_1\cdot k }{q_1\cdot k} +\frac{k^{\mu}q_2\cdot \epsilon^* -
\epsilon^{*\mu}q_2\cdot k }{q_2\cdot k} \right). 
\end{eqnarray}
Formally, $H_0^\mu(p,q_1,q_2)$ is as in the Born case, but with $p=q_1+q_2+k$
instead of $p=q_1+q_2$ for the
virtual photon propagator. Let us note, that the first Born-like term and 
two other terms in the second line of (\ref{eq:bremsAmp0}) are separately gauge invariant. 
The normalization of space-like part of $H_0$ is not as at  the
Born level: $|\vec q_1-\vec q_2| < \sqrt{s-4m_\pi^2}$. The time-like part of $H_0$ drops out when the product with $V_\mu$ is taken.  

Let us now return  to  $V_{\mu}$.
Following conventions of \cite{Jadach:1990zf} it reads: 
\begin{equation}\label{eq:born1}
V^\mu = 2\left (|\lambda_+|\hat{e}_1^\mu + i \lambda_+ \hat{e}_2^\mu
-m \lambda_- \hat{e}_3^\mu\right) \, .
\end{equation}
where $\lambda _{\pm}=\lambda_1\pm\lambda_2$. The vectors satisfy
$(\hat{e}_\alpha)^\mu = \delta^{\mu}_{\alpha}$.
We choose $\hat e_1$ to lie in the  reaction plane, while $\hat e_2 = p_1 \times
(q_1-q_2) / |p_1 \times (q_1-q_2) |$ is chosen to be perpendicular to that 
plane. The $\hat e_1$ is along incoming electron beam  and $\hat e_0$ is 
proportional to $p_1+p_2$. 
The basis vectors can be written as
\begin{equation}\label{eq:born2}
\hat{e}_0 = \left (\begin{array}{cccc}
1 & 0 &0 &0 \end{array}\right )\, , \hat{e}_1 = \left (\begin{array}{cccc}
0 & 1 &0 &0 \end{array}\right )\, , \hat{e}_2 = \left (\begin{array}{cccc}
0 & 0 &1 &0 \end{array}\right )\, ,  \hat{e}_3 = \left (\begin{array}{cccc}
0 & 0 &0 &1 \end{array}\right )\, .
\end{equation}
We can drop the term proportional to the electron mass. 

At Born level the second term in the expression (\ref{eq:born1}) will not contribute because $e_2\cdot
(q_1-q_2)=0$. The complete amplitude is thus:
\begin{equation}
M_{Born} = e^2 F_{2\pi}(S)\frac{1}{\sqrt{S}}|\lambda_+|\hat{e}_1\cdot (q_1-q_2),
\end{equation}
where $\sqrt{S}$ is the energy of c.m.. 
One can see that the amplitude is proportional to $\sin \theta_B $ as it should be.
Here $\theta_B = \angle p_1 q_1 $ is a scattering angle. Squared and summed 
over initial spin states the amplitude yelds:
\begin{equation}
\sum_{\lambda} |M_{Born}|^2 (S,T,U) = \frac{8 (4\pi \alpha)^2
 F_{2\pi}^2(S)}{S^2} 
\left (T U - m_\pi^2 S\right ).
\label{BornXX}
\end{equation}
The Mandelstam variables are defined as follows:
\begin{align}
S = 2 p_1\cdot p_2, \ \ \ T = 2 p_1 \cdot q_1, \ \ \ U = 2 p_1\cdot q_2.
\end{align}

The amplitude (\ref{eq:bremsAmp0}) for single photon emission can be 
decomposed into a sum of two gauge invariant parts:
\begin{equation}
H^\mu =H^\mu_{I} +H^\mu_{II}
\label{AmpliS}
\end{equation}
or
\begin{equation}
H^\mu =H^\mu_{I'} +H^\mu_{II'}
\label{AmplipS}
\end{equation}

where 
\begin{equation}
{H_I^\mu} =\frac{e^2F_{2\pi}(p^2)}{p^2}(q_1-q_2)^{\mu}
\left (\frac{q_1\cdot \epsilon^*}{q_1\cdot k} - 
\frac{q_2\cdot \epsilon^*}{q_2\cdot k}\right ), 
\label{HI}
\end{equation}
\begin{equation}
{H_{II}^\mu} = \frac{e^2F_{2\pi}(p^2)}{p^2}
\left (k^{\mu} \left (\frac{q_1\cdot \epsilon^*}{q_1\cdot k} +
\frac{q_2\cdot \epsilon^*}{q_2\cdot k}\right )-2 \epsilon^{*\mu}\right ), 
\label{HII}
\end{equation}

and alternatively 
\begin{equation}
H_{I'}^\mu =\frac{e^2F_{2\pi}(p^2)}{p^2}\left ((q_1-q_2)^{\mu}+k^{\mu}
\frac{q_2\cdot k - q_1\cdot k}{q_2\cdot k + q_1\cdot k} \right )
\left (\frac{q_1\cdot \epsilon^*}{q_1\cdot k} - 
\frac{q_2\cdot \epsilon^*}{q_2\cdot k}\right ), 
\label{HIp}
\end{equation}
\begin{eqnarray}
H_{II'}^{\mu} &=&  \frac{2e^2F_{2\pi}(p^2)}{p^2}
\left (\frac{k^{\mu}}{q_2\cdot k + q_1\cdot k}
(q_1\cdot \epsilon^*+ q_2\cdot \epsilon^*) - \epsilon^{*\mu}\right )\, .
\label{HIIp}
\end{eqnarray}

\section{Cross section and  amplitude separation}
Before going into numerical results, let us elaborate on the formulas
presented above in more details. One can see rather easily that formulas
(\ref{HI}) and (\ref{HIp}) have a form typical for amplitudes
of QED exclusive exponentiation \cite{Jadach:1998jb}, that is,  Born factors 
multiplied by an eikonal factor $\left (\frac{q_1\cdot \epsilon^*}{q_1\cdot k} - 
\frac{q_2\cdot \epsilon^*}{q_2\cdot k}\right )$. In fact, the two
expressions differ in the way how the Born factor approach
the genuine Born expression. In both cases the expressions  
approach the Born  in soft photon limit. In case of  (\ref{HIp})
this property holds for the photon  collinear to $\pi^+$ or $\pi^-$.
This was achieved by adding to (\ref{HIp}) the term proportional to $k^{\mu}
\frac{q_2\cdot k - q_1\cdot k}{q_2\cdot k + q_1\cdot k} $
and subtracting it from  (\ref{HIIp}). As a consequence the expression 
in the first bracket of (\ref{HIp}), in collinear configurations will be close to
$q_1-q_2\mp k$ respectively if $q_1 \cdot k \ll q_2 \cdot k$ and 
$q_2 \cdot k \ll q_1 \cdot k$. Thus, it is consistent with LL level factorization into
Born amplitude and eikonal factor. Generally expressions (\ref{HI}) and
(\ref{HIp}) differ from a product of Born times eikonal factor only by 
normalization. This defect is easy to correct, 
and we will return to this point later in this section when discussion of cross section will be given. 

Experience with the $Z\to l^+ l^-$ decay has shown that it is useful not only 
to rely on spin amplitudes, but to collect expressions for amplitudes squared 
and (partly) averaged over the spin degrees of freedom, since it can be useful
for future work on matching kernels of consecutive emissions.

If one takes separation (\ref{AmpliS}) for the calculation of two parts
of spin amplitudes, after spin average, an expression for the 
cross section based on (\ref{AmpliS}) takes the form:
\begin{eqnarray}
\sum_{\lambda,\epsilon} |M|^2 & = & \sum_{\lambda,\epsilon} |M_{I}|^2
+ \sum_{\lambda,\epsilon} |M_{II}|^2 + 2\sum_{\lambda,\epsilon} M_{I}
M_{II}^*\, ,
\label{Mone}
\end{eqnarray}
 where
\begin{eqnarray}
\sum_{\lambda,\epsilon} |M_I|^2 & = &- 2e^6 \frac{F_{2\pi}^2(S)}{S^2}
\left (\frac{m_\pi^2}{(q_1\cdot k)^2} + \frac{m_\pi^2}{(q_2\cdot
 k)^2}-  \frac {S^\prime}{(q_1\cdot k)(q_2\cdot k)}\right )
 \nonumber \\
&& \left \{\left (TU^\prime - m_\pi^2 S\right) +   
\left (T^\prime U - m_\pi^2 S\right) - 
\left (TT^\prime + UU^\prime - SS^\prime \right)
\right \} \nonumber \\
&& = -(\pi\alpha)\left (\frac{m_\pi^2}{(q_1\cdot k)^2} + 
\frac{m_\pi^2}{(q_2\cdot k)^2}-  
\frac {S^\prime}{(q_1\cdot k)(q_2\cdot k)}\right )
 \left ( A +  B + C+D \right )+ \nonumber \\
 && 8e^6 \frac{F_{2\pi}^2(S)}{S^2}
\left (\frac{m_\pi^2}{(q_1\cdot k)^2} + \frac{m_\pi^2}{(q_2\cdot
 k)^2}-  \frac {S^\prime}{(q_1\cdot k)(q_2\cdot k)}\right )
(q_1\cdot k)(q_2\cdot k),
\end{eqnarray}
\begin{eqnarray}
\sum_{\lambda,\epsilon} |M_{II}|^2 & = & - 2e^6
 \frac{F_{2\pi}^2(S)}{S^2}
\left\{\left (\frac{m_\pi^2}{(q_1\cdot k)^2} + \frac{m_\pi^2}{(q_2\cdot
 k)^2}+ \frac {S^\prime}{(q_1\cdot k)(q_2\cdot k)}\right ) \right.
\nonumber \\ 
&&\left. \left (\left (TU^\prime - m_\pi^2 S\right) +   
\left (T^\prime U - m_\pi^2 S\right) + 
\left (TT^\prime + UU^\prime - SS^\prime \right)\right)-8 S
 \right. \nonumber \\
&& \left. + \frac{4}{q_1\cdot k}\left (
TU^\prime - m_\pi^2 S + \frac{1}{2}
(TT^\prime + UU^\prime - SS^\prime) \right)
 \right. \nonumber \\
&& \left.+ \frac{4}{q_2\cdot k}\left (
T^\prime U - m_\pi^2 S + \frac{1}{2}
(TT^\prime + UU^\prime - SS^\prime) \right)
\right\}\nonumber \\
& = &  -(\pi\alpha) \frac{F_{2\pi}^2(S)}{S^2}
\left\{\left (\frac{m_\pi^2}{(q_1\cdot k)^2} + \frac{m_\pi^2}{(q_2\cdot
 k)^2}+ \frac {S^\prime}{(q_1\cdot k)(q_2\cdot k)}\right ) \right.
\nonumber \\ 
&&\left. (A +B -C -D) + \frac{4}{q_1\cdot k}\left(B-\frac{1}{2}(C+D)\right ) +
 \frac{4}{q_2\cdot k}\left(A-\frac{1}{2}(C+D)\right )\right \}\nonumber \\
&& - 8e^6 \frac{F_{2\pi}^2(S)}{S^2}
\left (\frac{m_\pi^2}{(q_1\cdot k)^2} + \frac{m_\pi^2}{(q_2\cdot
 k)^2}+  \frac {S^\prime}{(q_1\cdot k)(q_2\cdot k)}\right )
(q_1\cdot k)(q_2\cdot k)\nonumber \\
&& + 8e^6 \frac{F_{2\pi}^2(S)}{S^2}(S^\prime+ 2m_\pi^2 + S),
\end{eqnarray}
\begin{eqnarray}
2 \sum_{\lambda,\epsilon}M_{I} {M_{II}}^* & = & - 4e^6
 \frac{F_{2\pi}^2(S)}{S^2}
\left\{\left (\frac{m_\pi^2}{(q_1\cdot k)^2} - \frac{m_\pi^2}{(q_2\cdot
 k)^2}\right ) \right.
\nonumber \\ 
&&\left. \left (- \left (TU^\prime - m_\pi^2 S\right) +   
\left (T^\prime U - m_\pi^2 S\right)\right)
 \right. \nonumber \\
&& \left. - \frac{2}{q_1\cdot k}\left (
TU^\prime - m_\pi^2 S - \frac{1}{2}
(TT^\prime + UU^\prime - SS^\prime) \right)
 \right. \nonumber \\
&& \left.+ \frac{2}{q_2\cdot k}\left (
-(T^\prime U - m_\pi^2 S) + \frac{1}{2}
(TT^\prime + UU^\prime - SS^\prime) \right)
\right\} \nonumber \\
&& = -(\pi\alpha)\left\{
\left (\frac{m_\pi^2}{(q_1\cdot k)^2} -
 \frac{m_\pi^2}{(q_2\cdot k)^2}\right )
( 2A-2B)- \frac{4}{q_1\cdot k}\left (B + \frac{1}{2}(C +D)\right )
\right. \nonumber \\
&& \left.- \frac{4}{q_2\cdot k}\left( A + \frac{1}{2}(C +D)\right) 
\right \}+ 8e^6 \frac{F_{2\pi}^2(S)}{S^2}(S^\prime+ 2m_\pi^2 -S).
\end{eqnarray}
The definitions of terms $A, B, C, D, E$ will be given later in the section.

If instead of expression (\ref{AmpliS}) we use (\ref{AmplipS}) the following 
relation is obtained:

\begin{eqnarray}
\sum_{\lambda,\epsilon} |M|^2 & = & \sum_{\lambda,\epsilon} |M_{I'}|^2
+ \sum_{\lambda,\epsilon} |M_{II'}|^2 + 2\sum_{\lambda,\epsilon} M_{I'}
M_{II'}^*\, ,
\label{Mtwo}
\end{eqnarray}
where
\begin{eqnarray}
\sum_{\lambda,\epsilon} |M_{I'}|^2 & = &- 8e^6 \frac{F_{2\pi}^2(S)}{S^2}
\left (\frac{m_\pi^2}{(q_1\cdot k)^2} + \frac{m_\pi^2}{(q_2\cdot
 k)^2}-  \frac {S^\prime}{(q_1\cdot k)(q_2\cdot k)}\right )
 \nonumber \\
&& \left \{ \frac {(q_1\cdot k)^2}{(q_1\cdot k + q_2\cdot k)^2}  
\left (TU^\prime - m_\pi^2 S\right) + 
\frac {(q_2\cdot k)^2}{(q_1\cdot k + q_2\cdot k)^2}  
\left (T^\prime U - m_\pi^2 S\right) - \right.\nonumber \\
&& \left. \frac {(q_1\cdot k)(q_2\cdot k)}{(q_1\cdot k + q_2\cdot k)^2}
\left (TT^\prime + UU^\prime - SS^\prime \right)
\right \} \nonumber \\
&& = -(4\pi\alpha)\left (\frac{m_\pi^2}{(q_1\cdot k)^2} + 
\frac{m_\pi^2}{(q_2\cdot k)^2}-  
\frac {S^\prime}{(q_1\cdot k)(q_2\cdot k)}\right )\nonumber \\
&& \left (\frac {(q_2\cdot k)^2}{(q_1\cdot k + q_2\cdot k)^2} A 
+ \frac {(q_1\cdot k)^2}{(q_1\cdot k + q_2\cdot k)^2} B 
+  \frac {(q_1\cdot k)(q_2\cdot k)}{(q_1\cdot k + q_2\cdot k)^2}
( C+D ) \right )+ \nonumber \\
 && 32e^6 \frac{F_{2\pi}^2(S)}{S^2}
\left (\frac{m_\pi^2}{(q_1\cdot k)^2} + \frac{m_\pi^2}{(q_2\cdot
 k)^2}-  \frac {S^\prime}{(q_1\cdot k)(q_2\cdot k)}\right )
\frac{(q_1\cdot k)^2(q_2\cdot k)^2}{(q_1\cdot k + q_2\cdot k)^2}\; '
\end{eqnarray}
\begin{eqnarray}
\sum_{\lambda,\epsilon} |M_{II'}|^2 & = &\frac{- 8e^6F_{2\pi}^2(S)}{S^2}
\left [\frac{S}{(q_1\cdot k + q_2\cdot k)^2} 
\left (TU^\prime - m_\pi^2 S + T^\prime U - m_\pi^2 S + 
TT^\prime + UU^\prime - SS^\prime  \right)\right. \nonumber \\
&& \left. - 2S\right]\nonumber \\
& = & \frac{- (4\pi\alpha)S}{(q_1\cdot k + q_2\cdot k)^2}
( A +B - C- D)+ \frac{16e^6F_{2\pi}^2(S)}{S^2}
\frac{(q_1\cdot k)^2 + (q_2\cdot k)^2}{(q_1\cdot k +
 q_2\cdot k)^2}S\, . 
 \end{eqnarray}
Note that $\sum_{\lambda,\epsilon} |M_{II'}|^2$ is free of infrared and collinear divergences. 
The interference contribution is given by the following expression:
\begin{eqnarray}
 2\sum_{\lambda,\epsilon} M_{I'} M_{II'}^*& =&\frac{- 8e^6F_{2\pi}^2(S)}{S^2}
\frac{1}{(q_1\cdot k + q_2\cdot k)^2}\left \{ \left ((2m_\pi^2 + S^\prime)
\frac{q_1\cdot k}{q_2\cdot k} - S\right)\left (TU^\prime - m_\pi^2 S
 \right) + \right. \nonumber \\
&& \left. \left ((2m_\pi^2 + S^\prime)
\frac{q_2\cdot k}{q_1\cdot k} - S\right)\left (T^\prime U - m_\pi^2 S
\right)\right. + \nonumber \\
&& \left. \left (\frac{S}{2}\left ( \frac{q_1\cdot k}{q_2\cdot k} +
 \frac{q_2\cdot k}{q_1\cdot k}\right ) - 2m_\pi^2 - S^\prime \right ) 
\left (TT^\prime + UU^\prime - SS^\prime  \right)
  \right \}\nonumber \\
& = & \frac{-4\pi\alpha}{(q_1\cdot k + q_2\cdot k)^2}
\left \{ \left ((2m_\pi^2 + S^\prime)
\frac{q_2\cdot k}{q_1\cdot k} - S\right)A + \left ((2m_\pi^2 + S^\prime)
\frac{q_1\cdot k}{q_2\cdot k} - S\right)B \right. \nonumber \\
&& \left. -\left (\frac{S}{2}\left ( \frac{q_1\cdot k}{q_2\cdot k} +
 \frac{q_2\cdot k}{q_1\cdot k}\right ) - 2m_\pi^2 - S^\prime \right
 )(C+D)\right \} - \nonumber \\
&& \frac{32e^6F_{2\pi}^2(S)}{S^2(q_1\cdot k + q_2\cdot k)^2}
\left [\frac{S}{2}\left ((q_1\cdot k)^2 +
 (q_2\cdot k)^2 \right )
- (2m_\pi^2 + S^\prime)(q_1\cdot k)(q_2\cdot k) \right ].\nonumber \\ && 
\end{eqnarray}

The Mandelstam variables are defined as follows
\begin{align}
S = 2p_1 \cdot p_2, \ \ \ S^\prime = 2q_1 \cdot q_2,  \\
T = 2p_1 \cdot q_1, \ \ \ T^\prime = 2 p_2 \cdot q_2,  \\
U = 2p_1\cdot q_2, \ \ \ U^\prime = 2 p_2 \cdot q_1.  
\end{align}
Finally
\begin{align}
A = \sum_{\lambda} |M_{Born}|^2(S, T^\prime,U), \\
B = \sum_{\lambda}|M_{Born}|^2(S,T, U^\prime), \\
C = \sum_{\lambda}|M_{Born}|^2(S, T, U), \\
D = \sum_{\lambda}|M_{Born}|^2(S, T^\prime,U^\prime), \\
E =  32(4 \pi \alpha)^3 m_\pi^2\frac{F_{2\pi}^2(S)}{S^2}. 
\end{align}
Let us point that the  complete expression for the amplitude squared
is, in comparison to its parts, short:
\begin{equation}
\sum_{\lambda,\epsilon} |M|^2 = 
4 \pi \alpha \left \{\frac{-m_\pi^2}{(q_1\cdot k)^2} A +
 \frac{-m_\pi^2}{(q_2\cdot k)^2}B + 
\frac{S-2m_\pi^2}{2(q_1\cdot k)(q_2\cdot k)}
\left (C +D \right )\right \} +E  \, .
\label{elegant}
\end{equation}

We should stress that our two separation options (eqs.(\ref{Mone}) and (\ref{Mtwo}))
can have their first terms even closer to  Born-times-eikonal-factor 
form. For that purpose it is enough to adjust normalization of (\ref{HI}) 
(or (\ref{HIp})) to Born amplitude times eikonal factor.
 Compensating adjustment to 
(\ref{HII}) (or (\ref{HIIp})) is then necessary. 
The changes can be performed by numerical manipulation of the three
 contributions to (\ref{Mone}) and (\ref{Mtwo}). 
The resulting new separation into parts will be distinguished by additional prime over its parts.
For example  $ \sum_{\lambda,\epsilon} |M_{II}^\prime|^2$ will be used instead of   
$ \sum_{\lambda,\epsilon} |M_{II}|^2$.

Such a modification  is of interest, because if  $\sum_{\lambda,\epsilon} |M_{I}^\prime|^2$ 
or $ \sum_{\lambda,\epsilon} |M_{I'}^\prime|^2$ is used alone,
then it is the expression used in simulation with {\tt PHOTOS} Monte Carlo and
 refinement of 
 \cite{Nanava:2006vv}. 
In the next section,
we will perform our numerical investigations with 
respect to results obtained from formulas of ref. \cite{Nanava:2006vv} 
(which for our present process is just an  approximation). 

\section{General numerical results}
We have performed our numerical studies  for the decaying photon virtualities of
2, 20, 200 and 2000 GeV. However in this paper we will show only the case of 
2 GeV. The other ones confirm only that the collinear logarithms are properly 
reproduced by the simulation with standard set-up of the {\tt PHOTOS} kernel and would 
not add anything relevant to our discussion.

Let us start with a presentation of the case when the weight for the matrix element
is that for $ \sum_{\lambda,\epsilon} |M_{I}^\prime|^2$ or $ \sum_{\lambda,\epsilon} |M_{I'}^\prime|^2$. 
As one can see from fig.\ref{fig:one} agreement with matrix element of \cite{Nanava:2006vv} is excellent over
all  the phase space. Unfortunately, as it will be discussed in the 
next section,  this is true only for the case when distributions are averaged 
over the orientation of the whole event with respect to incoming beams 
(or spin state of the virtual photon). At this moment let us note that as 
a consequence of strongly varying Born cross section (approaching zero in 
forward and backward direction) the resulting weight distribution from
  $\sum_{\lambda,\epsilon} |M_{I}^\prime|^2$ or 
$\sum_{\lambda,\epsilon} |M_{I'}^\prime|^2$ has a tail. We have used 
special techniques to appropriately adapt Monte Carlo simulation to that.

If $\sum_{\lambda,\epsilon} |M_{I}|^2$ or 
$ \sum_{\lambda,\epsilon} |M_{I'}|^2$ is used directly instead of $\sum_{\lambda,\epsilon} |M_{I}^\prime|^2$ or $ \sum_{\lambda,\epsilon} |M_{I'}^\prime|^2$,
 normalization of 
Born-like factor is not performed, differences with respect to formulas of
\cite{Nanava:2006vv} are much larger, see respectively figs \ref{fig:two}
and \ref{fig:three}. In the last case discrepancies are smaller, because the 
normalization is correct in collinear limit. Finally let us compare
results of complete scalar QED matrix element with that of \cite{Nanava:2006vv}, see fig.~\ref{fig:four}. 
In the high photon energy region there is a clear surplus of events with respect to the formulas of
\cite{Nanava:2006vv}. That contribution should not be understood as bremsstrahlung, but rather as a genuine process. 
Anyway in that region of phase space scalar QED is not expected to work well. 
Even though expression (\ref{elegant}) looks elegant and is short,
it needs to be separated into (at the cross section level) longer expressions, 
where Born times eikonal factor part is explicitly separated.
Note the difference between results shown on figs 
 \ref{fig:one} and  \ref{fig:four}  is only 0.2 \% of the total  
rate. That is why
our detailed discussion is not important for numerical conclusions,
but for the understanding 
of the underlying structure of distributions.
Once the status of approximations used in {\tt PHOTOS} at single photon 
radiation is understood, 
we can, as in other processes iterate and   
 to simulate effects of multibremsstrahlung
simultaneously with the detector effects. As an example we show in 
fig. \ref{fig:exp} results of the
single photon emission mode, and compare with the one of multiple emission mode. 
Differences are rather small. This may not be  the case if selection cuts are present.

Now, let us consider the decay $W\to l \nu(\gamma)$. In \cite{Nanava:2003cg} 
a simple correcting weight was introduced into {\tt PHOTOS}
for discrepancies with respect to exact predictions of {\tt SANC} \cite{Andonov:2004hi}.
A weight based on the exact matrix element is presently available, see Appendix.
One can check again how good  approximation of ref.~\cite{Nanava:2003cg} is.
As one can see from fig. \ref{Gizo1} the correction weight reproduces  the 
result of exact matrix
element well. In fig. \ref{Gizo2}, we show that once the exact matrix element is 
implemented into {\tt PHOTOS} the agreement with the benchmark calculation is better than the statistical error of $10^8$ events. 
In contrary to the previously studied $\gamma^* \to \pi^+\pi^- \gamma$ case,
there were no problems of   weight distribution tail.
\begin{figure*} 
\psfrag{M2pipi}{\scriptsize \bf $M_{\pi+\pi-}^2/S$}
\psfrag{M2PiPlusGamma}{\hspace*{-1mm}\scriptsize \bf 
$M_{\pi+\gamma}^2/S$}
\caption{\label{fig:one} \it \small Distributions of invariant masses  
normalized to center of mass energy and squared ($M^{\;2}/S$) for
$e^+e^- \to \pi^+\pi^- (\gamma)$ at 2 GeV center of mass energy. Results 
from  {\tt PHOTOS} with matrix element taken from \cite{Nanava:2006vv} are given in red  (or darker grey) colour. 
If matrix element $ \sum_{\lambda,\epsilon}|M_{I}^\prime|^2$ or $ \sum_{\lambda,\epsilon} |M_{I'}^\prime|^2$ is used 
(the two options are effectively identical) results
are given in green colour. Logarithmic scale is used, but 
for the ratio (black line) linear scale is used instead. 
Fraction of events with photons above
50 MeV is respectively 4.2279 $\pm$ 0.0021 \% and 4.2269 $\pm$ 0.0021\% for the two programs. } 
\begin{center}
\subfigure[This distribution is identical 
to the distribution of photon energy in the reaction frame as well.
 ]{\includegraphics[%
  width=0.48\columnwidth]{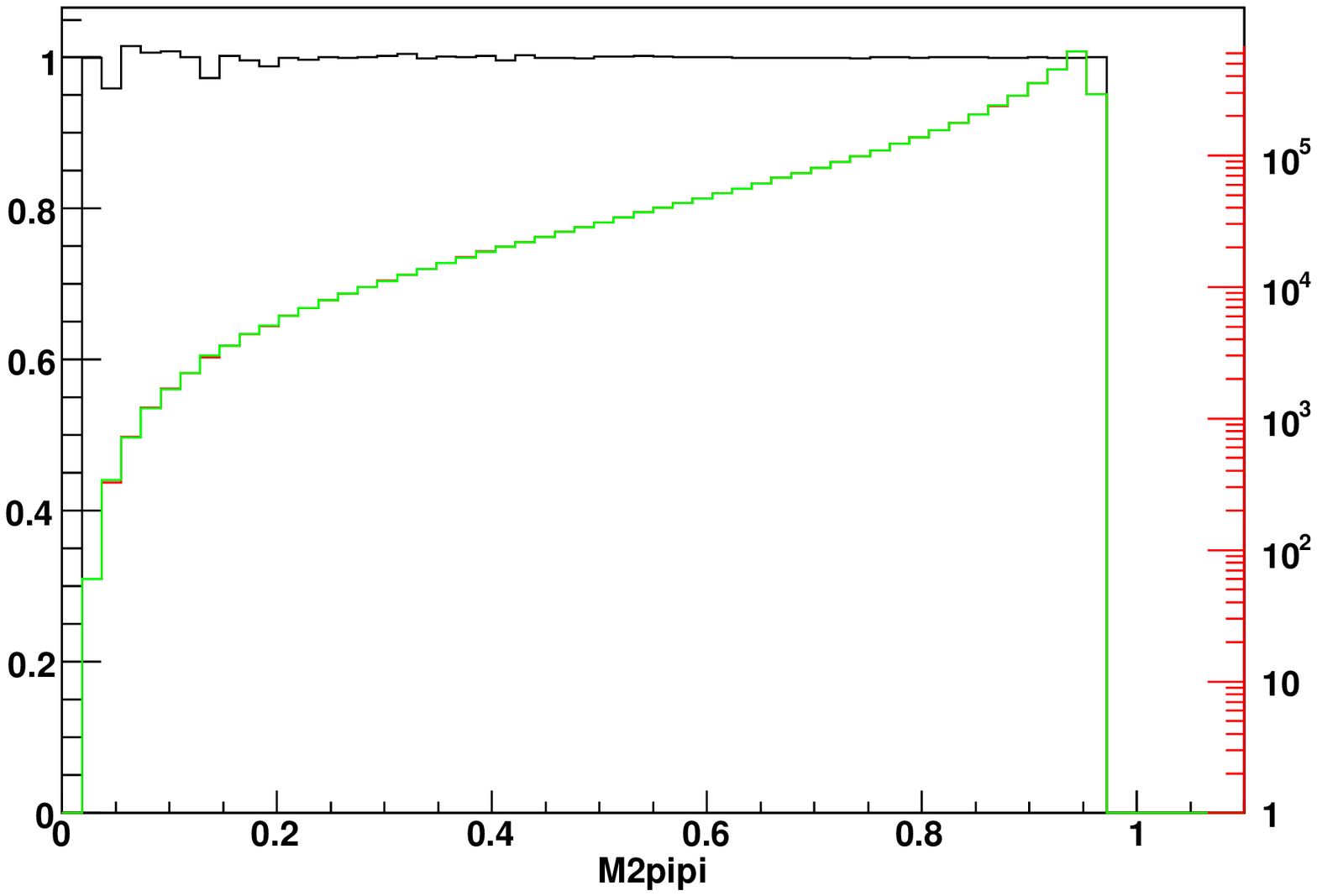}}
\hspace{3mm}
\subfigure[ {\it This distribution coincides with distributions for
 squared invariant mass  of $\pi^- \gamma$ pair.}
 ]{\includegraphics[%
 width=0.48\columnwidth]{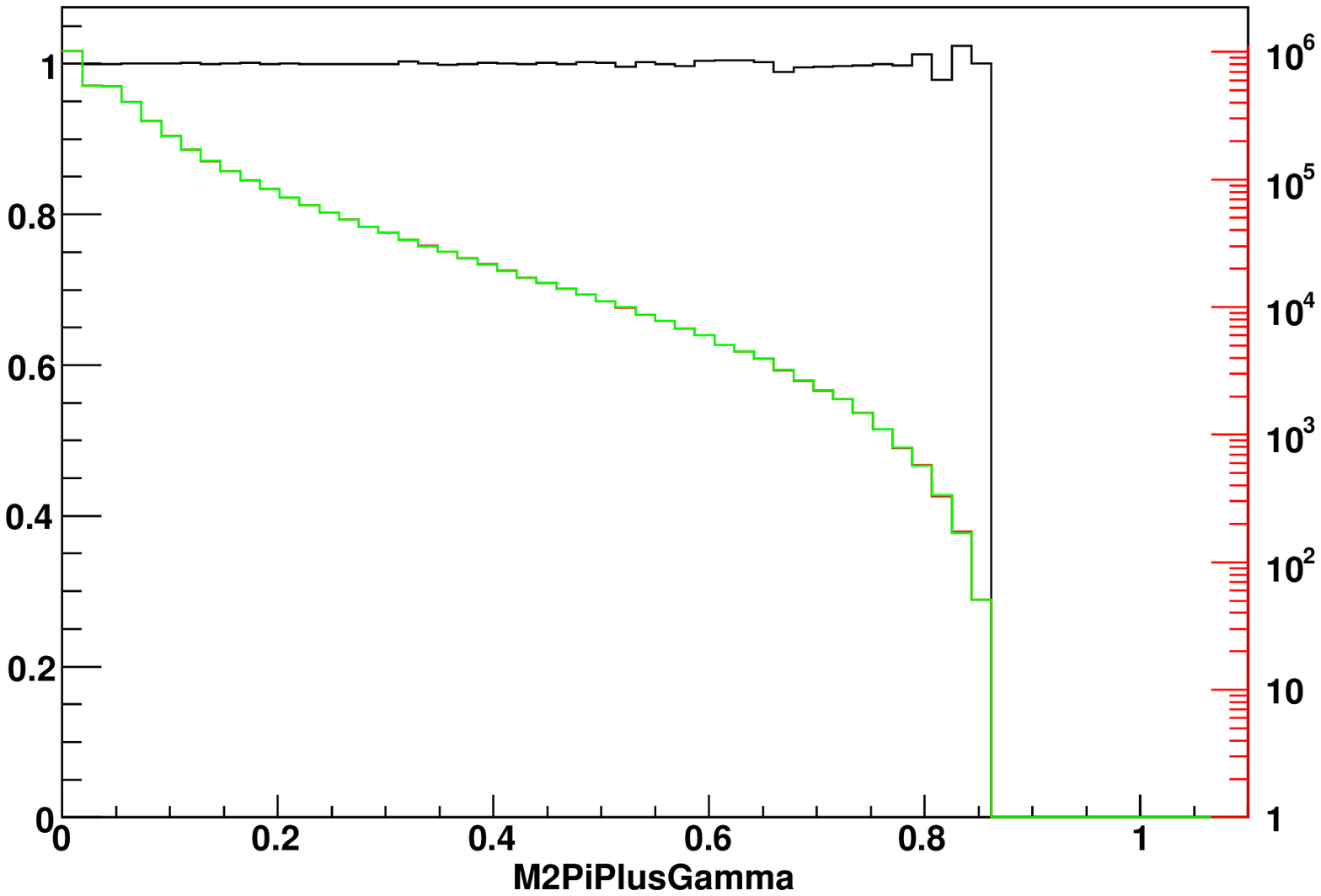}}
\end{center}
\end{figure*}

\begin{figure*}
\psfrag{M2pipi}{\scriptsize \bf $M_{\pi+\pi-}^2/S$}
\psfrag{M2PiPlusGamma}{\hspace*{-1mm}\scriptsize \bf $M_{\pi+\gamma}^2/S$}
\caption{\label{fig:two} \it \small Distributions of invariant masses  
normalized to center of mass energy and squared ($M^{\;2}/S$) for
$e^+e^- \to \pi^+\pi^- (\gamma)$ at 2 GeV center of mass energy. Results 
from  {\tt PHOTOS} with matrix element taken from \cite{Nanava:2006vv} are given in red colour. If matrix element $|M_{I}|^2$  is  used  results
are given in green colour. Logarithmic scale is used, but
for the ratio (black line) linear scale is used instead. 
Fraction of events with photons above
50 MeV is respectively 4.2279 $\pm$ 0.0021  \% and 3.4435  $\pm$ 0.0019 \% for the two programs.  
}

\begin{center}
\subfigure[This distribution is 
identical to the distribution of photon energy in the reaction frame as well.
 ]{\includegraphics[%
  width=0.48\columnwidth,
  keepaspectratio]{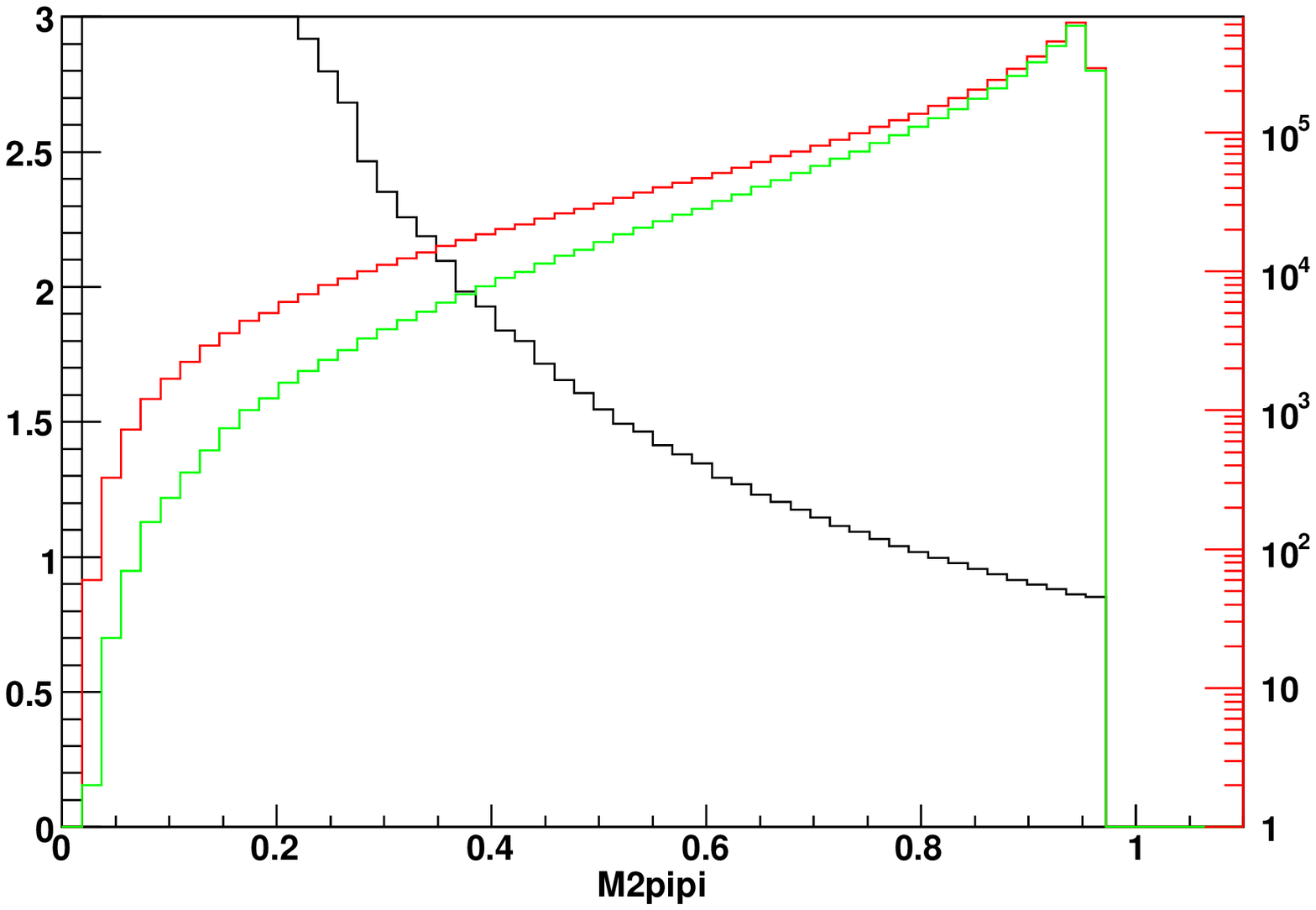}}
\hspace{3mm}
\subfigure[ {\it 
 It coincides with distributions for  squared invariant mass  of $\pi^- \gamma$ pair.}
 ]{\includegraphics[%
 width=0.48\columnwidth,
  keepaspectratio]{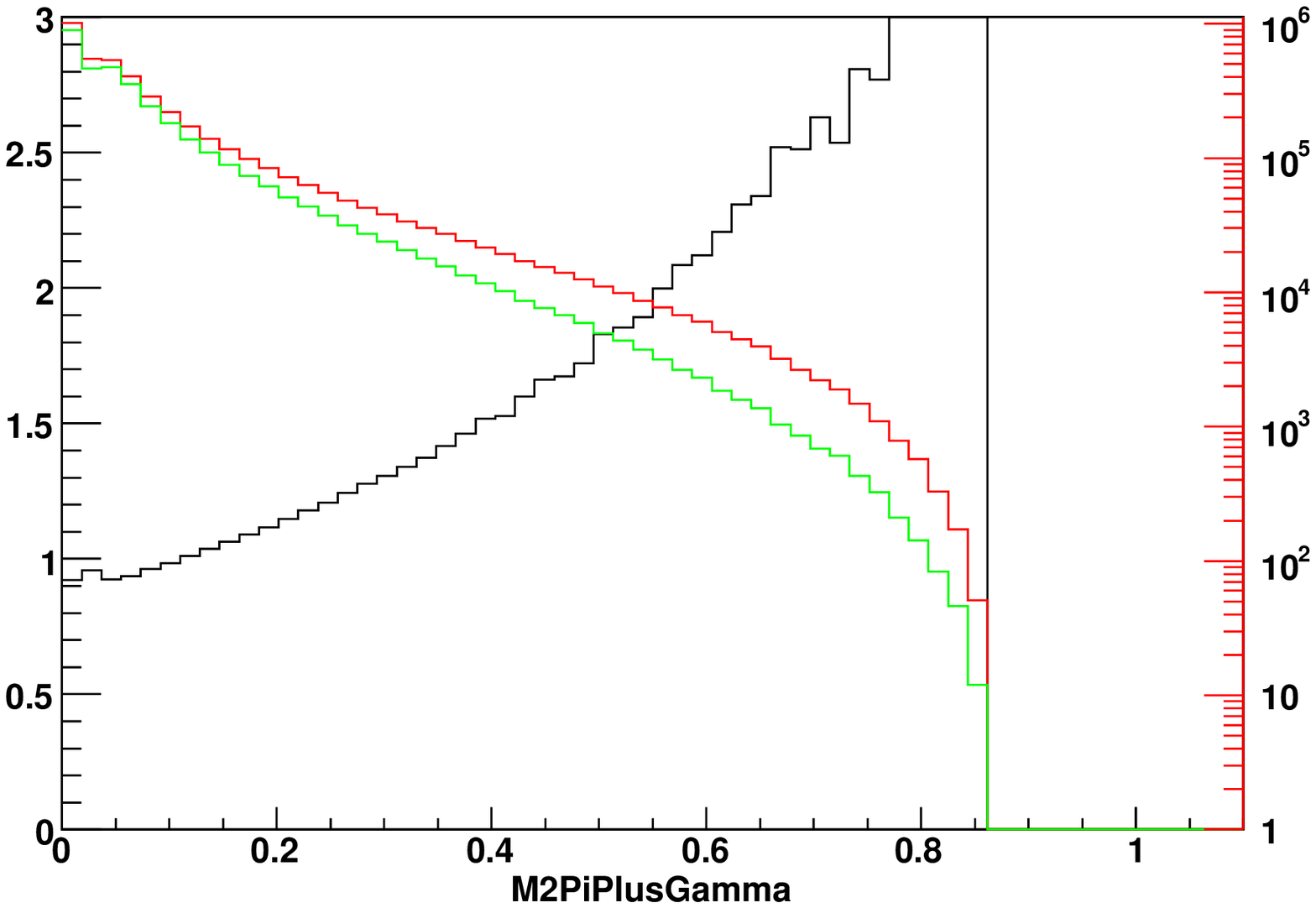}}
\end{center}
\end{figure*}

\begin{figure*}
\psfrag{M2pipi}{\scriptsize \bf $M_{\pi+\pi-}^2/S$}
\psfrag{M2PiPlusGamma}{\hspace*{-1mm}\scriptsize \bf $M_{\pi+\gamma}^2/S$}
\caption{\label{fig:three} \it \small Distributions of invariant masses  
normalized to center of mass energy and squared ($M^{\;2}/S$) for
$e^+e^- \to \pi^+\pi^- (\gamma)$ at 2 GeV center of mass energy. Results 
from  {\tt PHOTOS} with matrix element taken from \cite{Nanava:2006vv} are given in red colour. 
If matrix element $ \sum_{\lambda,\epsilon} |M_{I'}|^2$ is  used  results
are given in green colour. Logarithmic scale is used, but
for the ratio (black line) linear scale is used instead. 
Fraction of events with photons above
50 MeV is respectively 4.2279 $\pm$ 0.0021 \% and 3.8329  $\pm$ 0.0020 \% for the two programs.  
}

\begin{center}
\subfigure[This distribution is 
identical to the distribution of photon energy in the reaction frame as well.
 ]{\includegraphics[%
  width=0.48\columnwidth,
  keepaspectratio]{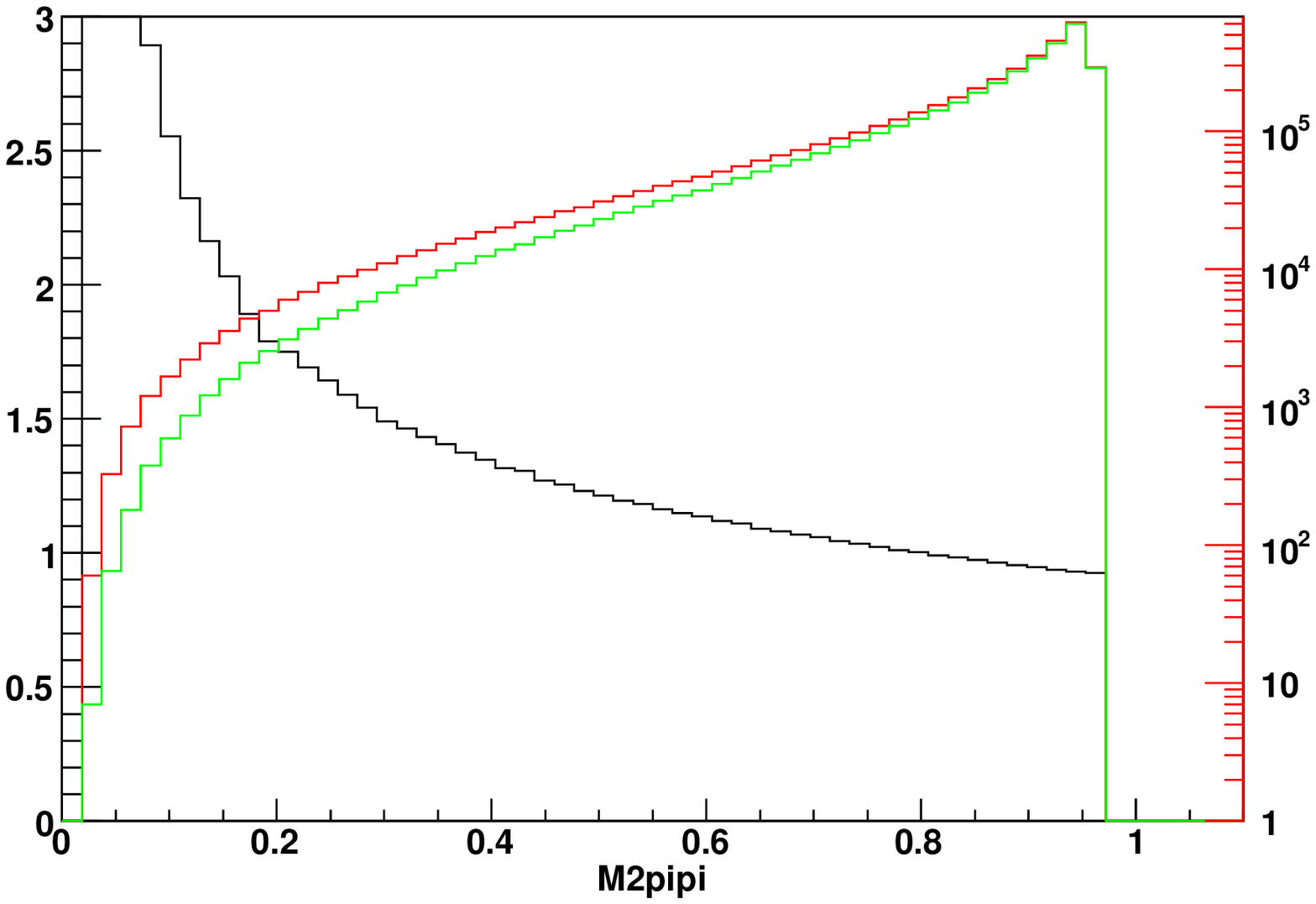}}
\hspace{3mm}
\subfigure[{\it It coincides with distributions for  squared invariant mass  
of $\pi^- \gamma$ pair.} ]{\includegraphics[%
 width=0.48\columnwidth,
  keepaspectratio]{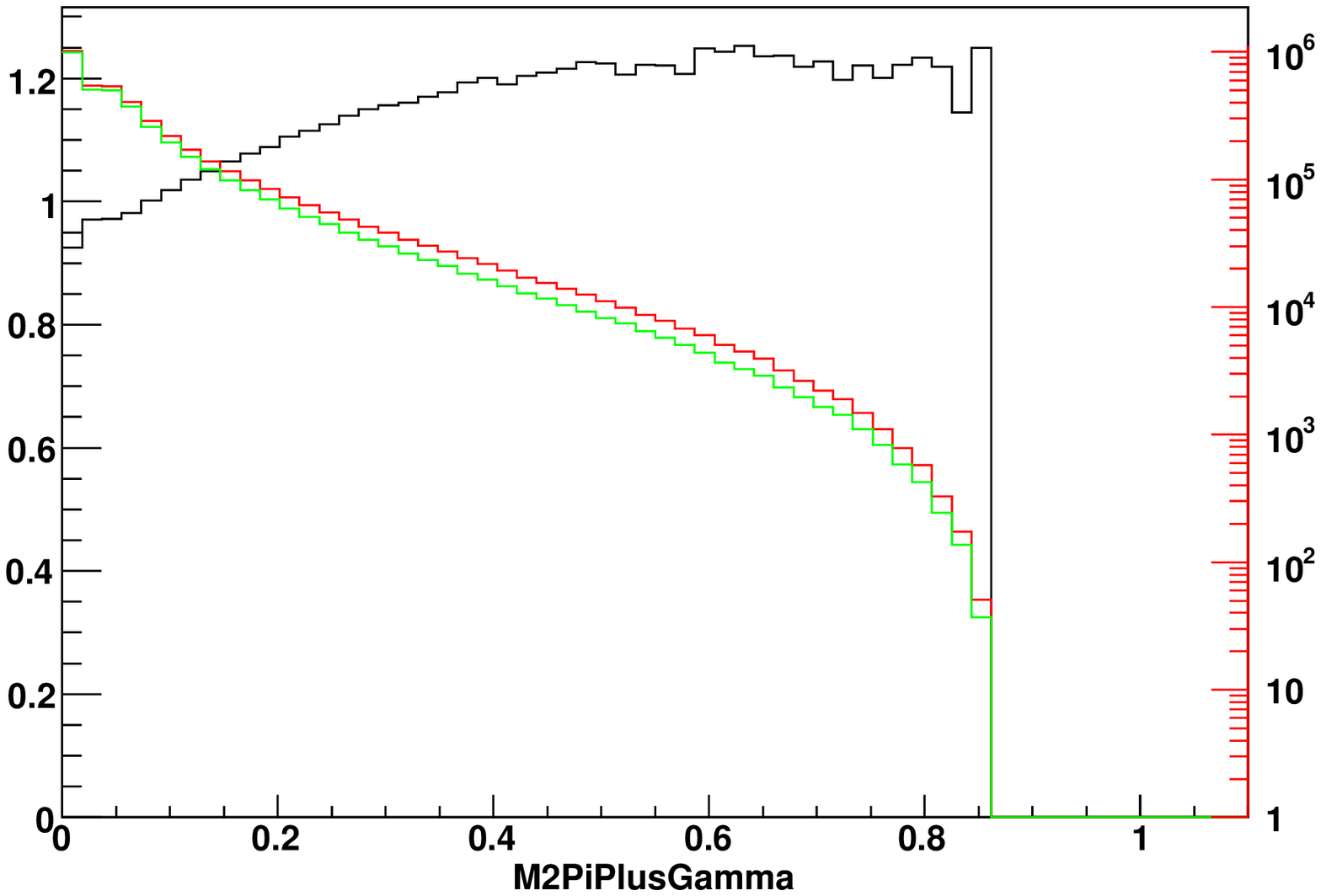}}
\end{center}
\end{figure*}

\begin{figure*}
\psfrag{M2pipi}{\scriptsize \bf $M_{\pi+\pi-}^2/S$}
\psfrag{M2PiPlusGamma}{\hspace*{-1mm}\scriptsize \bf $M_{\pi+\gamma}^2/S$}
\caption{\label{fig:four} \it \small Distributions of invariant masses  
normalized to center of mass energy and squared ($M^{\;2}/S$) for
$e^+e^- \to \pi^+\pi^- (\gamma)$ at 2 GeV center of mass energy. Results 
from  {\tt PHOTOS} with matrix element taken from \cite{Nanava:2006vv} are given in red colour. If complete matrix element is  used  results
are given in green colour. Logarithmic scale is used, but
for the ratio (black line) linear scale is used instead. 
Fraction of events with photons above
50 MeV is respectively  4.2279 $\pm$ 0.0021 \% and 4.4320 $\pm$ 0.0021\% for the two programs.  
}

\begin{center}
\subfigure[This distribution is 
identical to the distribution of photon energy in the reaction frame as well.
 ]{\includegraphics[%
  width=0.48\columnwidth,
  keepaspectratio]{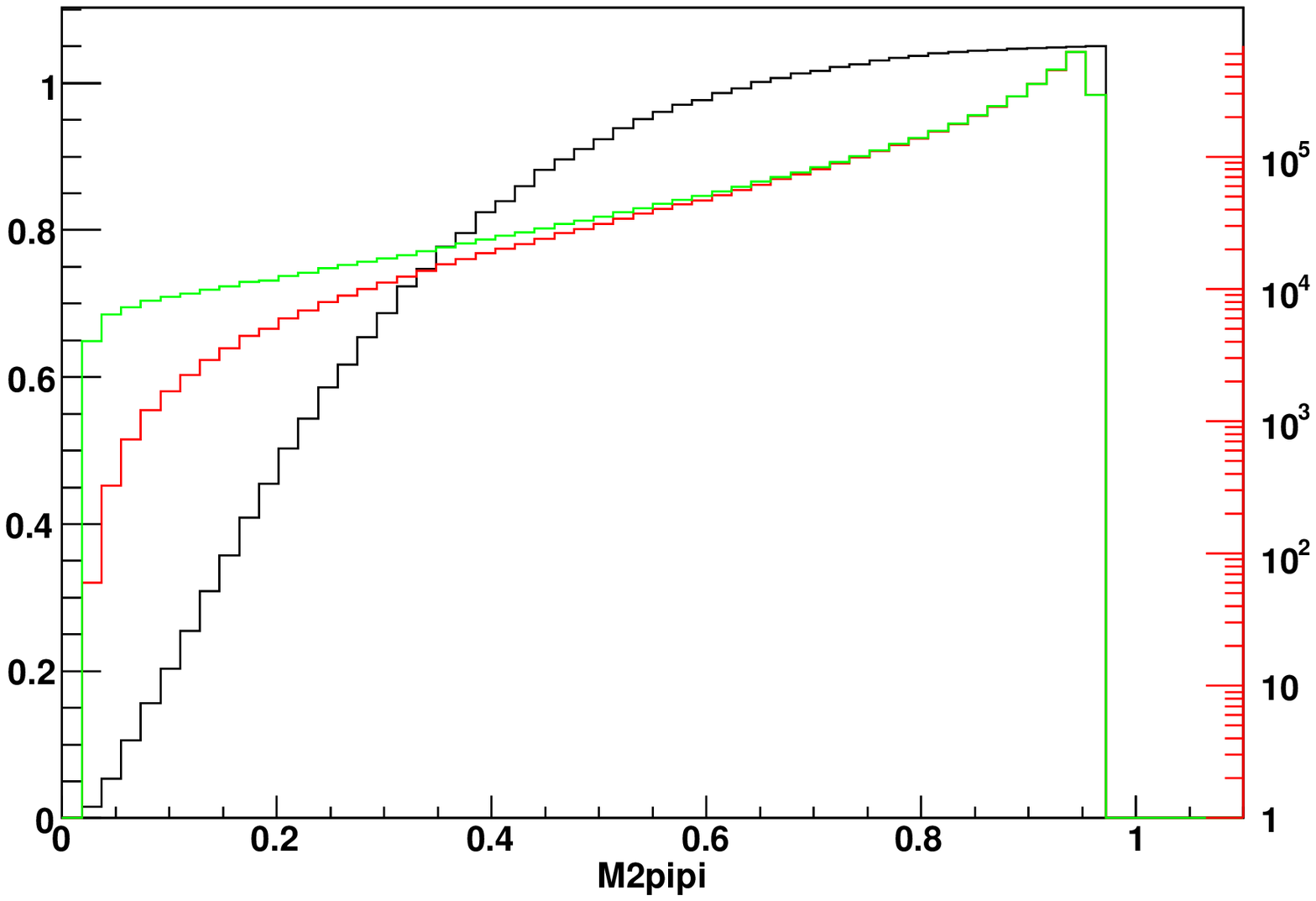}}
\hspace{3mm}
\subfigure[ {\it 
 It coincides with distributions for  squared invariant mass  of $\pi^- \gamma$ pair.}
 ]{\includegraphics[%
 width=0.48\columnwidth,
  keepaspectratio]{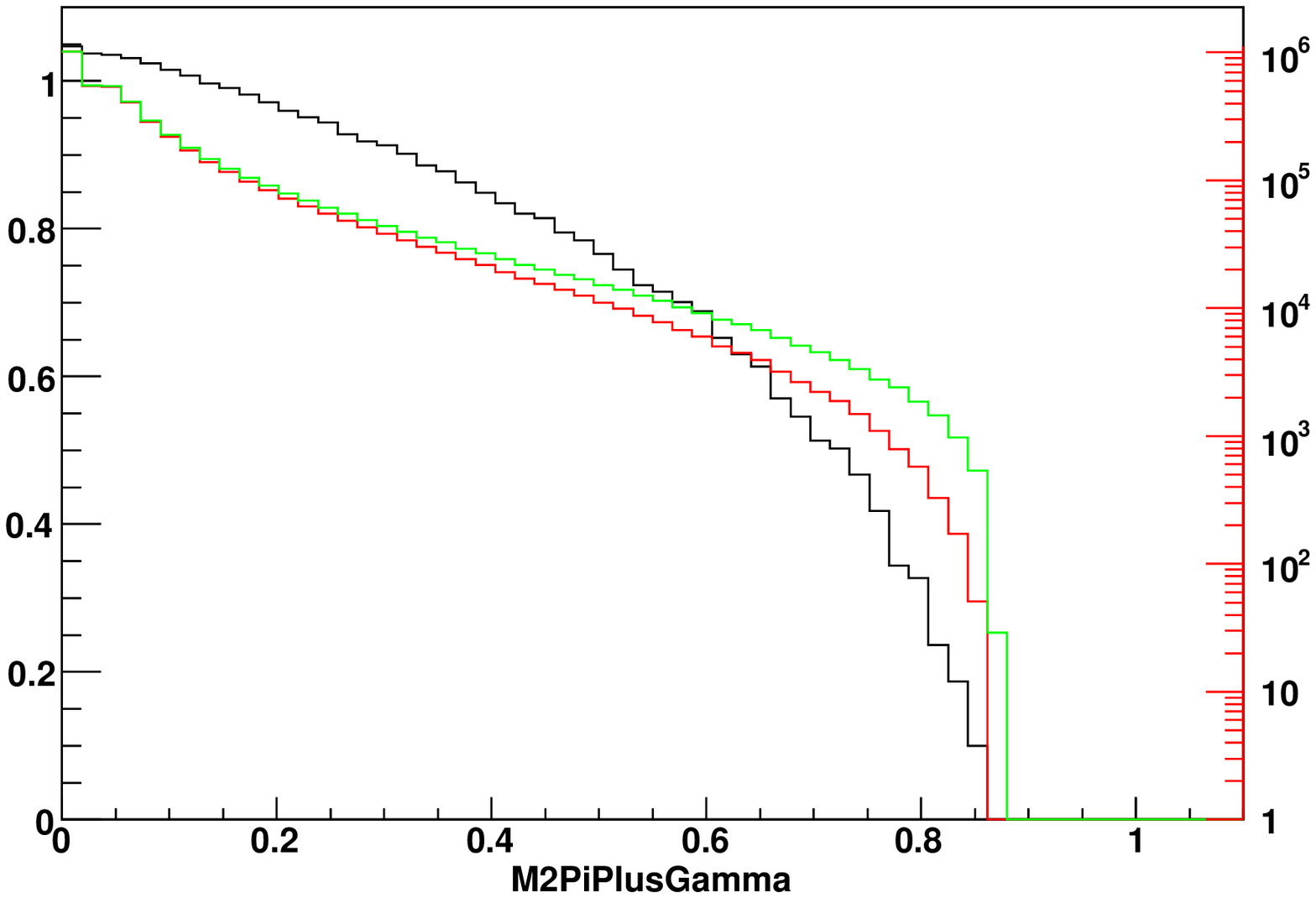}}
\end{center}
\end{figure*}

\begin{figure*}
\psfrag{M2pipi}{\scriptsize \bf $M_{\pi+\pi-}^2/S$}
\psfrag{M2PiPlusGamma}{\hspace*{-1mm}\scriptsize \bf $M_{\pi+\gamma}^2/S$}
\caption{\label{fig:exp} \it \small Distributions of invariant masses  
normalized to center of mass energy and squared ($M^{\;2}/S$) for
$e^+e^- \to \pi^+\pi^- (\gamma)$ at 2 GeV center of mass energy. Results 
from  {\tt PHOTOS} with matrix element taken from \cite{Nanava:2006vv} are
 given in red colour. If exponentiation is  activated as well 
 results
are given in green colour. Logarithmic scale is used, but
for the ratio (black line) linear scale is used instead. 
Fraction of events with
at least one  photon above
50 MeV is respectively  4.2279 $\pm$ 0.0021 \% and 4.1377 $\pm$ 0.0020\% for the two cases.  
}

\begin{center}
\subfigure[ This distribution 
is identical to the distribution of photon energy in the reaction frame as well. 
 ]{\includegraphics[%
  width=0.48\columnwidth,
  keepaspectratio]{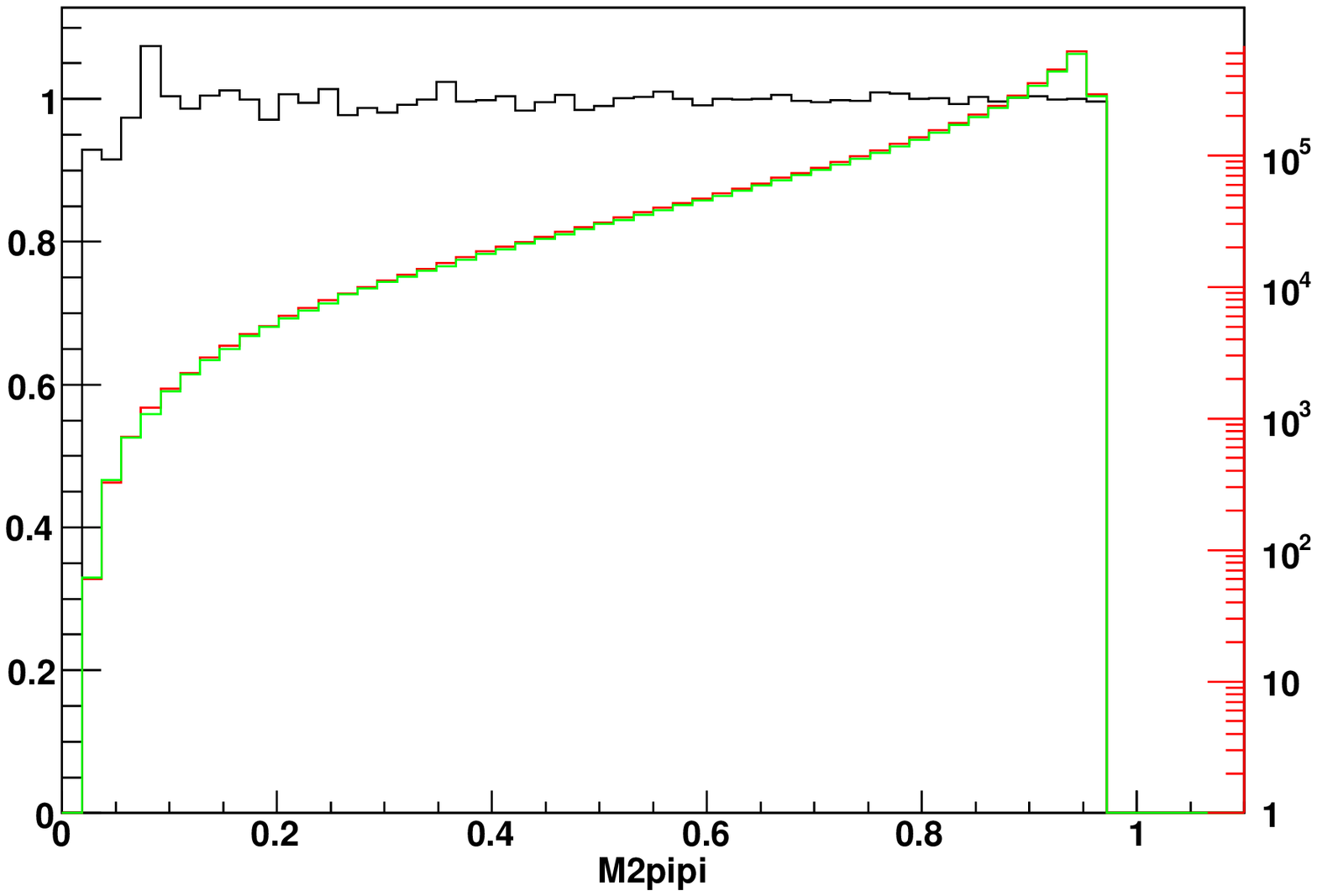}}
\hspace{3mm}
\subfigure[ {\it
 It coincides with distributions for  squared invariant mass  of $\pi^- \gamma$ pair. }
 ]{\includegraphics[%
 width=0.48\columnwidth,
  keepaspectratio]{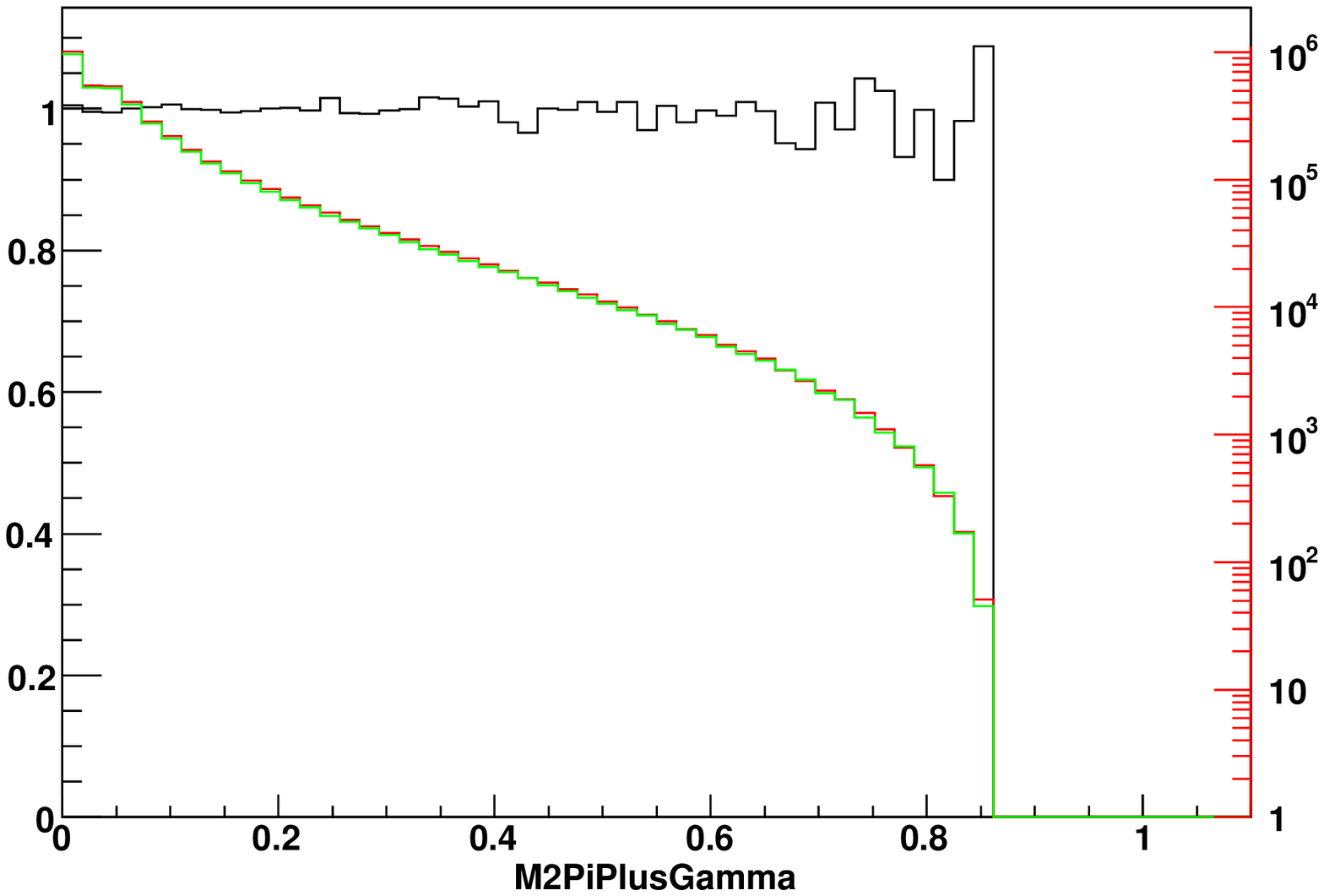}}
\end{center}
\end{figure*}

\begin{figure}[!h]
  \begin{center}
    \includegraphics[ width=180mm,height=120mm, keepaspectratio]{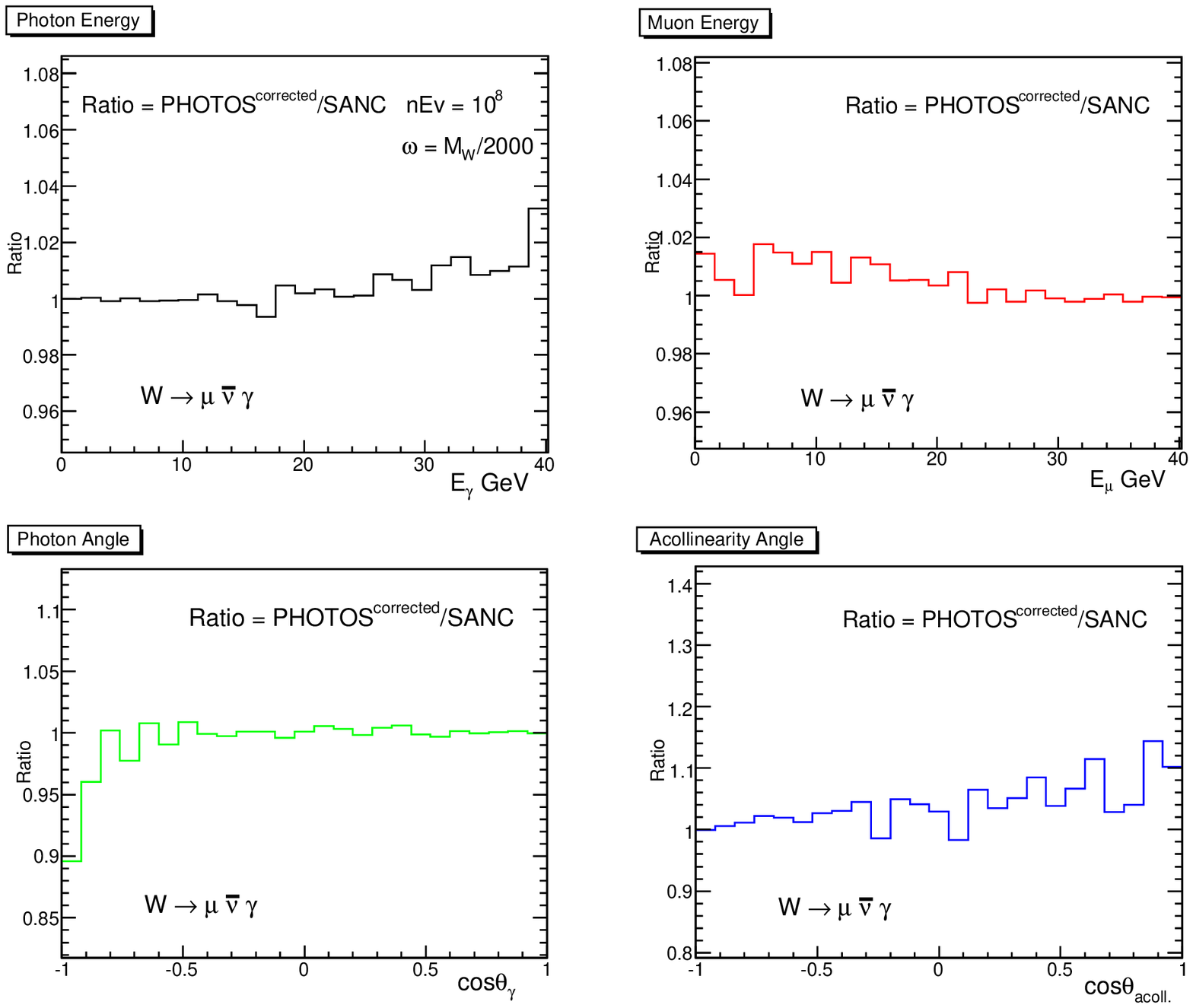}
  \end{center}
  \vspace*{-.8cm}
  \caption{\label{Gizo1} \it \small Ratios of  results from {\tt PHOTOS} with the old correcting weight and {\tt SANC} for  
    the $W\to \mu \bar{\nu} (\gamma)$ decay distributions. 
  }
\end{figure}

\begin{figure}[!h]
  \begin{center}
    \includegraphics[ width=180mm,height=120mm, keepaspectratio]{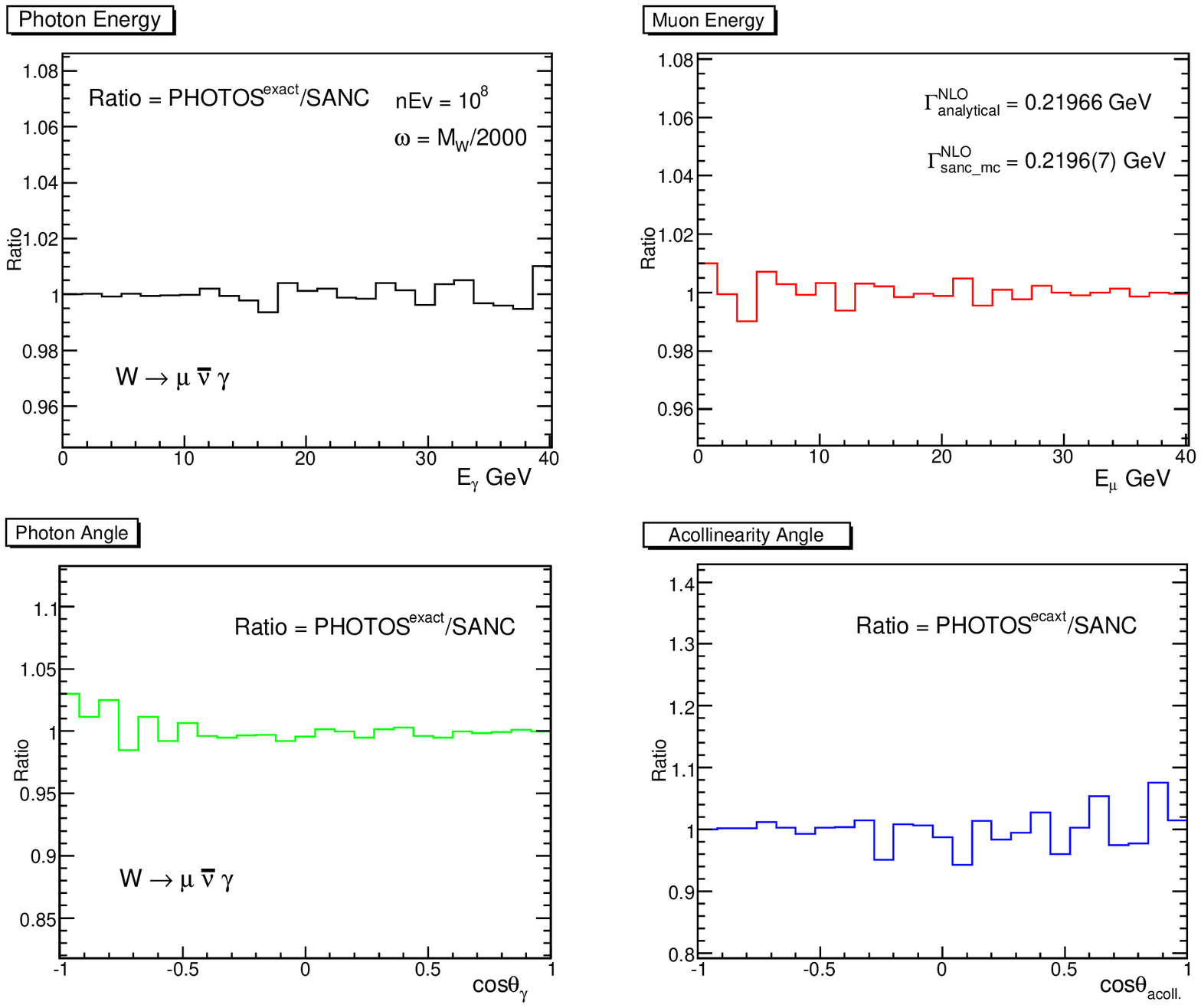}
  \end{center}
  \vspace*{-.8cm}
  \caption{\label{Gizo2} \it \small Ratios of results from {\tt PHOTOS} with the excact correcting weight and {\tt SANC} for  
    the $W\to \mu \bar{\nu} (\gamma)$ decay distributions. 
  }
\end{figure}

\section{Numerical results using beam direction}



In the previous section we have discussed distributions regarding four-momenta 
of decay products only. Agreement between results of {\tt PHOTOS} using
universal kernel and simulations based on matrix element was excellent
both in case of $W \to l \nu (\gamma) $ and   
$\gamma^* \to \pi^+ \pi^- (\gamma) $ decays, even though
 the decaying particle spin  effects were not taken into account\footnote{This is definitely
a complication requiring some attention.  It is an interesting aspect of the  
validation of {\tt PHOTOS}, absent in  the scalar state \cite{Nanava:2006vv}, but 
present in case of $Z$ decay \cite{Golonka:2006tw}, and it is strongly related
to limits of factorization.}  in the {\tt PHOTOS} kernel. 

As can be seen from  plots \ref{Gizo3}--\ref{Gizo3p2}, 
distributions in variables sensitive to the  orientation of the $W$ boson spin 
 are affected. The plots \ref{Gizo3} and \ref{Gizo4} show distributions
in the photon momentum angle with respect to 
a spin quantization axis as predicted by {\tt SANC} and by {\tt PHOTOS} with the standard kernel in transversally and
longitudinally polarized W boson decays. 
The plots \ref{Gizo3p1} and \ref{Gizo3p2} correspond to the muon momentum 
orientation.
The regions of phase space, where 
distributions are sparcely populated and where in fact at Born level probability 
density approach zero, are becoming moderately overpopulated by {\tt PHOTOS}
(increase of up to 14 \% of density was found for transversely polarized $W$ boson decay). In most cases, this is probably
 of no practical consequences, nonetheless it requires quantification.  
Once the exact matrix element is 
implemented into {\tt PHOTOS}, agreement with the {\tt SANC} predictions is better than statistical error of $10^8$ events,
see fig.~\ref{Gizo5}.

Similar effects take place
 for  $\gamma^* \to \pi^+\pi^- \gamma$. Even though 
from fig. \ref{fig:one} one could conclude that the universal kernel of \cite{Nanava:2006vv},
for arbitrary large samples,
is equivalent to the matrix element as given by
 $ \sum_{\lambda,\epsilon} |M_{I}^\prime|^2$ or
 $ \sum_{\lambda,\epsilon} |M_{I'}^\prime|^2$,      
differences appear in distributions sensitive to initial state spin orientation, 
see figs \ref{fig:oneX} and \ref{fig:oneY}. On these plots angular
distributions of the photon momentum with respect to the beam line are shown.
Again, regions of phase space giving zero contribution at the Born level
are becoming overpopulated if an approximation for the photon radiation matrix element is used. 
From that perspective and 
for practical reasons one can conclude that 
the $ \sum_{\lambda,\epsilon} |M_{I'}^\prime|^2$ choice is  better than 
$ \sum_{\lambda,\epsilon} |M_{I}^\prime|^2$.
It yelds distributions closer 
to the ones obtained from universal kernel. 
Then, the remaining part of (\ref{Mtwo}) represents better  
correction to implement bulk of effects from complete matrix element
to events generated with default  {\tt PHOTOS}.
We can see also that  distributions obtained from  
 kernel of ref. \cite{Nanava:2006vv}  and $ \sum_{\lambda,\epsilon} |M_{I}^\prime|^2$ are  close
 to each other but nonetheless distinct. Complete implementation requires
control of the $\gamma^*$ spin.
The  Born level distribution for $\gamma^*$ decay has zero at 
$\cos\theta_{\pi^\pm}=\pm 1$. Close to these directions
internal weight of {\tt PHOTOS} necessary for exact matrix element 
becomes large, see figs.~\ref{fig:oneX},\ref{fig:oneY}. 
Zero which is also present
in distribution of Born level $W$ decay  is of no such consequences for the
weight distribution.

\begin{figure}[!h]
  \begin{center}
      \includegraphics[ height=85mm, keepaspectratio]{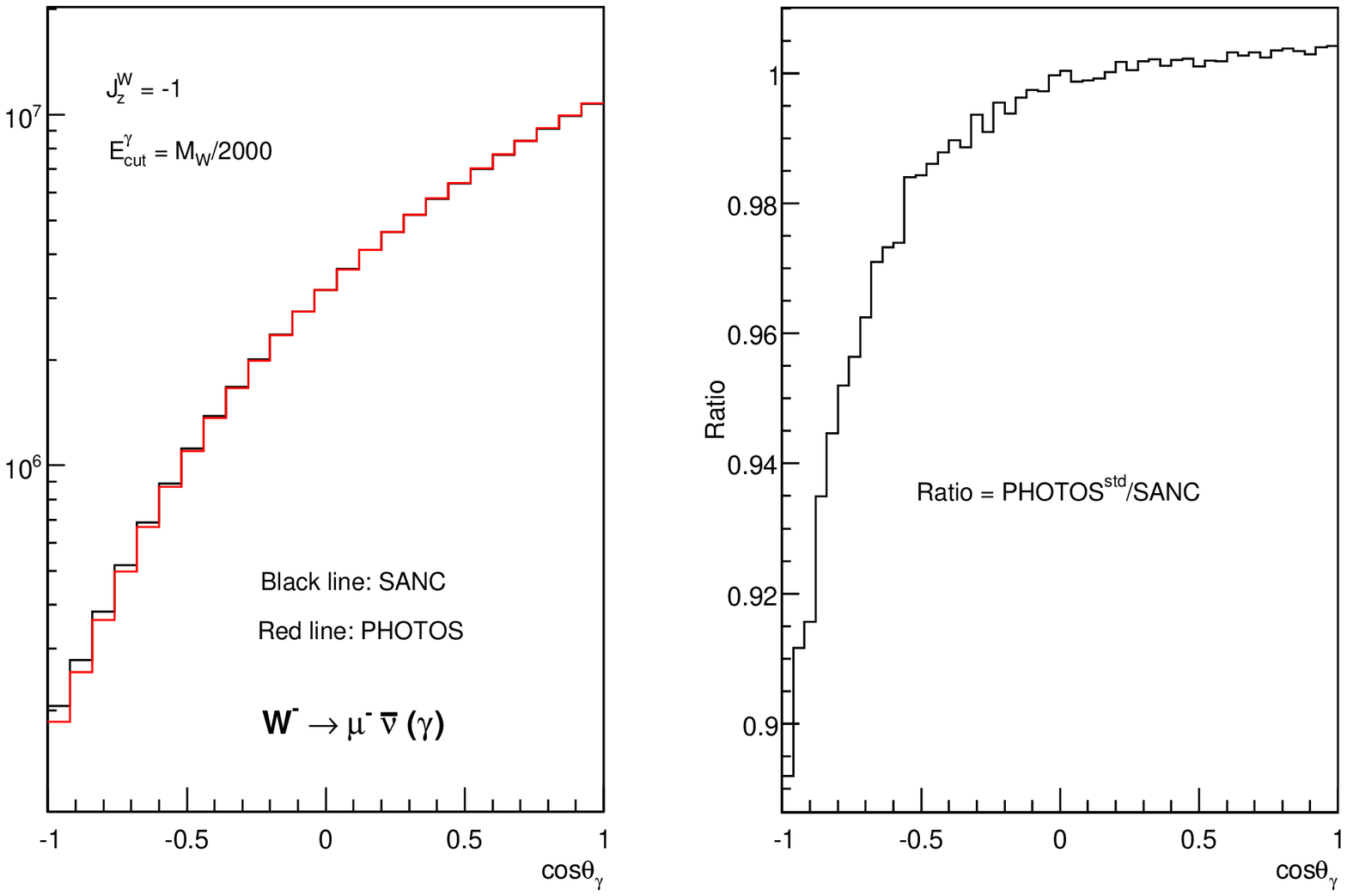}
  \end{center}   
  \vspace*{-1.cm}
  \caption{\label{Gizo3} \it \small Cosine of the angle of photon momentum with respect to spin axis in the decay  $W\to \mu \bar{\nu} (\gamma)$. 
   W boson is transversely polarized, $J^{W}_{z} = -1$. The black line corresponds to {\tt SANC} results,
   while the red line corresponds to predictions of {\tt PHOTOS} with standard kernel.
  }
\end{figure}

\begin{figure}[!h]
  \begin{center}
      \includegraphics[height=80mm, keepaspectratio]{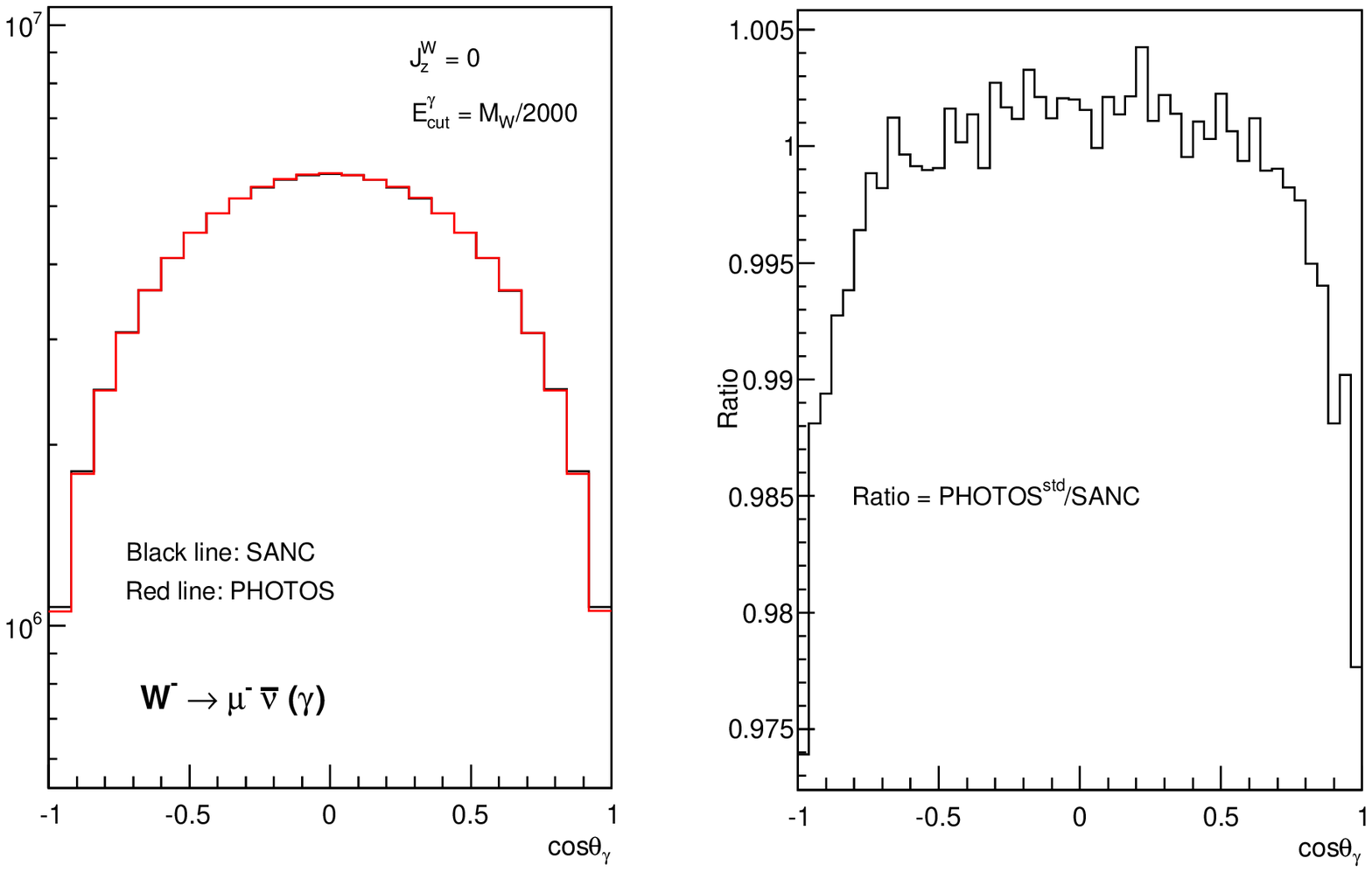}
  \end{center}
  \vspace*{-1.cm}
  \caption{\label{Gizo4} \it \small Cosine of the angle of photon momentum with respect to spin axis in the decay  $W\to \mu \bar{\nu} (\gamma)$. 
   W boson is longitudinally polarized, $J^{W}_{z} = 0$. The black line corresponds to {\tt SANC} results,
   while the red line corresponds to predictions of {\tt PHOTOS} with standard kernel.
  }
\end{figure}

\begin{figure}[!h]
  \begin{center}
    \includegraphics[height=83mm, keepaspectratio]{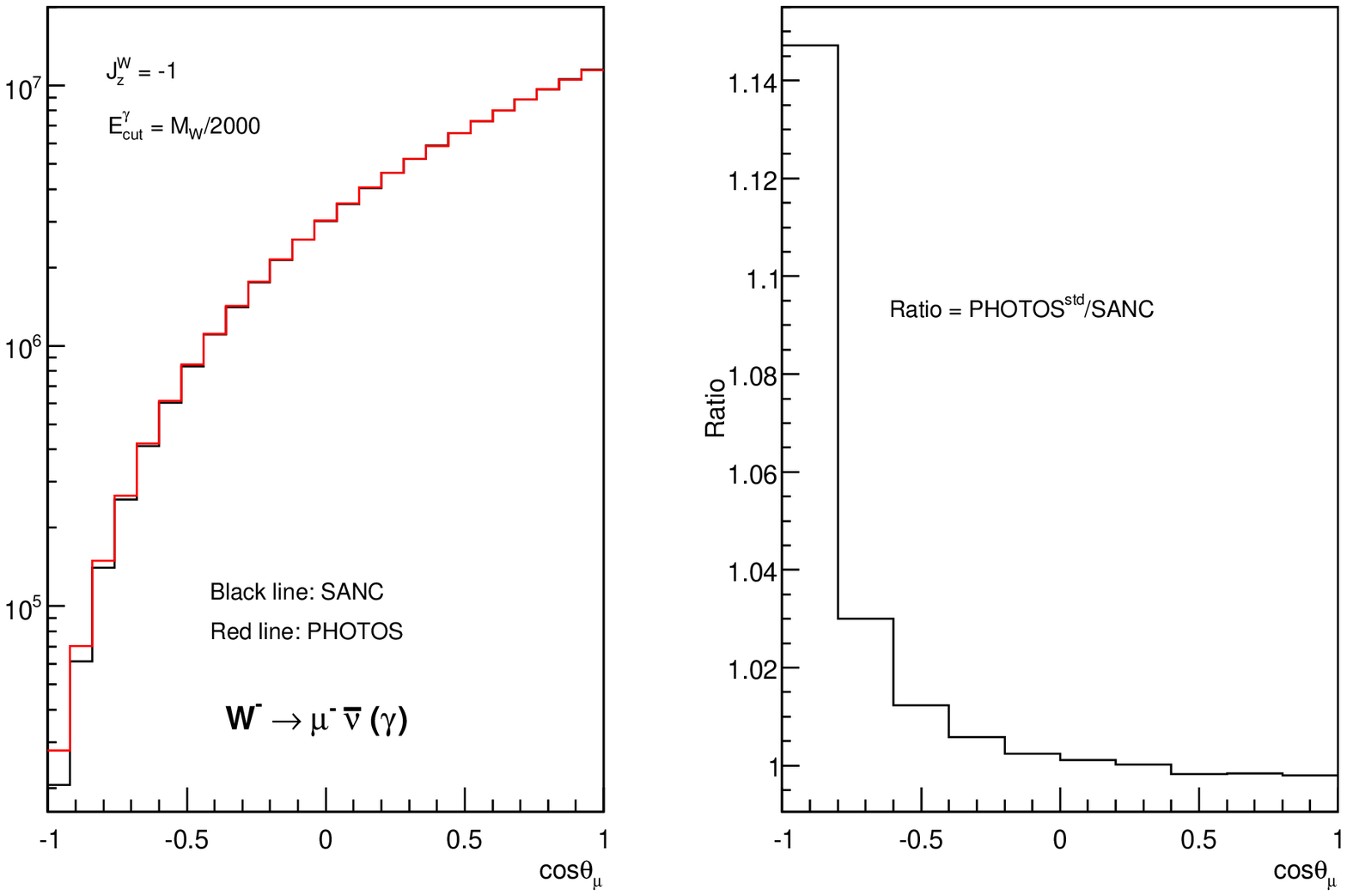}
  \end{center}
  \vspace*{-1.cm}
  \caption{\label{Gizo3p1} \it \small   Cosine of the angle of muon momentum with respect to spin axis in the decay  $W\to \mu \bar{\nu} (\gamma)$ is shown.
   W boson is transversally polarized, $J^{W}_{z} = -1$. The black line corresponds to {\tt SANC} results,
   while the red line corresponds to predictions of {\tt PHOTOS} with standard kernel.
Comparison of the distributions show that 
    lack of spin effects in {\tt PHOTOS} standard kernel results 
in up to 14\% surplus of events 
for    phase-space region close to $\cos\theta_\mu=-1$. 
  }
\end{figure}

\begin{figure}[!h]
  \begin{center}
    \includegraphics[height=79mm, keepaspectratio]{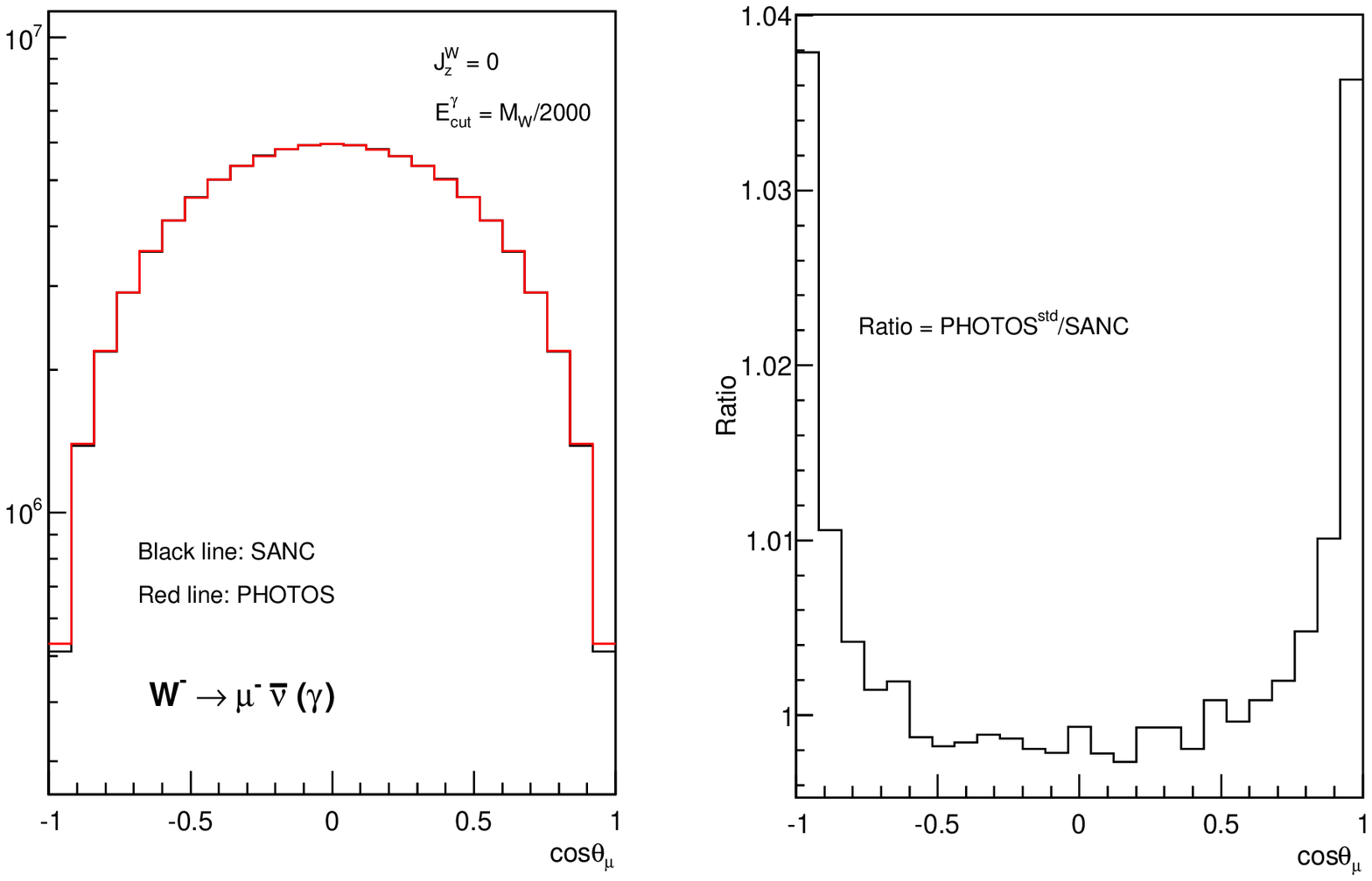}
  \end{center}
  \vspace*{-1.cm}
  \caption{\label{Gizo3p2} \it \small   Cosine of the angle of the muon momentum with respect to spin axis in the decay  $W\to \mu \bar{\nu} (\gamma)$ is shown. 
   W boson is longitudinally polarized, $J^{W}_{z} = 0$. The black line corresponds to {\tt SANC} results,
   while the red line corresponds to predictions of {\tt PHOTOS} with standard kernel.
Comparison of the distributions show that 
    lack of spin effects in {\tt PHOTOS} standard kernel results 
in up to 4\% surplus of events  
for    phase-space region close to $\cos\theta_\mu=\pm 1$. 
  }\end{figure}

\begin{figure}[!h]
  \begin{center}
      \includegraphics[ height=92mm, keepaspectratio]{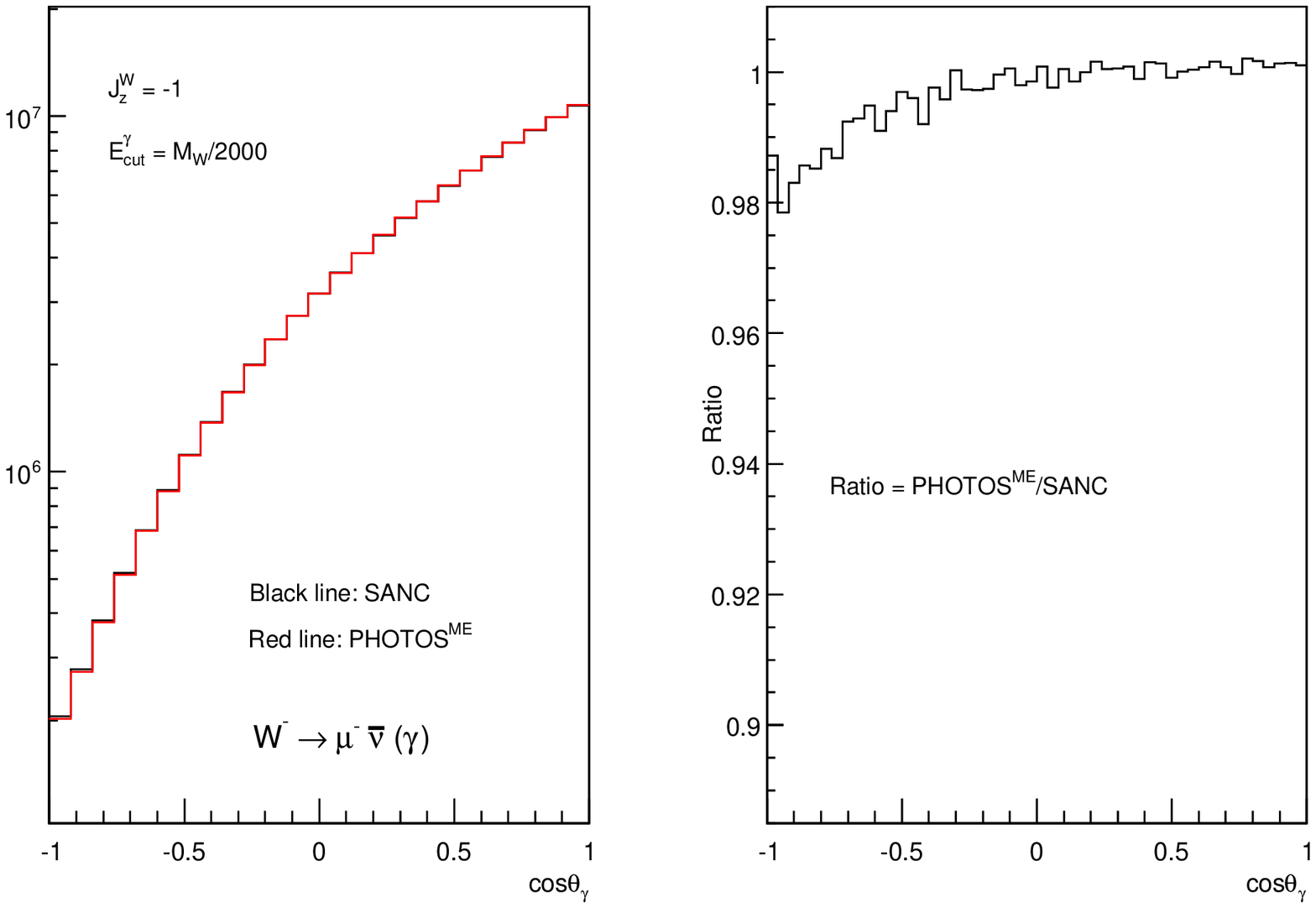}
  \end{center}
  \vspace*{-1.cm}
  \caption{\label{Gizo5} \it \small Cosine of the angle of photon momentum with respect to spin axis in the decay  $W\to \mu \bar{\nu} (\gamma)$ is shown. 
   W boson is transversely polarized, $J^{W}_{z} = -1$, along this axis. The black line corresponds to {\tt SANC} results,
   while the red line corresponds to predictions of {\tt PHOTOS} with exact matrix element.
  }
\end{figure}

\begin{figure*}[!h]
\psfrag{cosThetaGamma}{\scriptsize $\cos\theta_{\gamma}$}
\psfrag{cosThetaPiPlus}{\scriptsize $\cos\theta_{\pi+}$}
\caption{\label{fig:oneX} \it \small Angular distributions  for
$e^+e^- \to \pi^+\pi^- (\gamma)$ at 2 GeV center of mass energy. Results 
from  {\tt PHOTOS} with matrix element taken from \cite{Nanava:2006vv} are given in red colour. Matrix element $\sum_{\lambda,\epsilon} |M_{I}^\prime|^2$ 
is  used for results with green line. Logarithmic scale is used, but
for the ratio (black line) linear scale is used instead. 
 Fraction of presented events (i.e. with photons above
50 MeV) is respectively 4.2279 $\pm$ 0.0021 \% and 4.2269 $\pm$ 0.0021\% 
of the total samples for the two programs.  
}

\begin{center}
\subfigure[cosine of the photon angle with respect to beam. 
 ]{\includegraphics[%
  width=0.49\columnwidth,
  keepaspectratio]{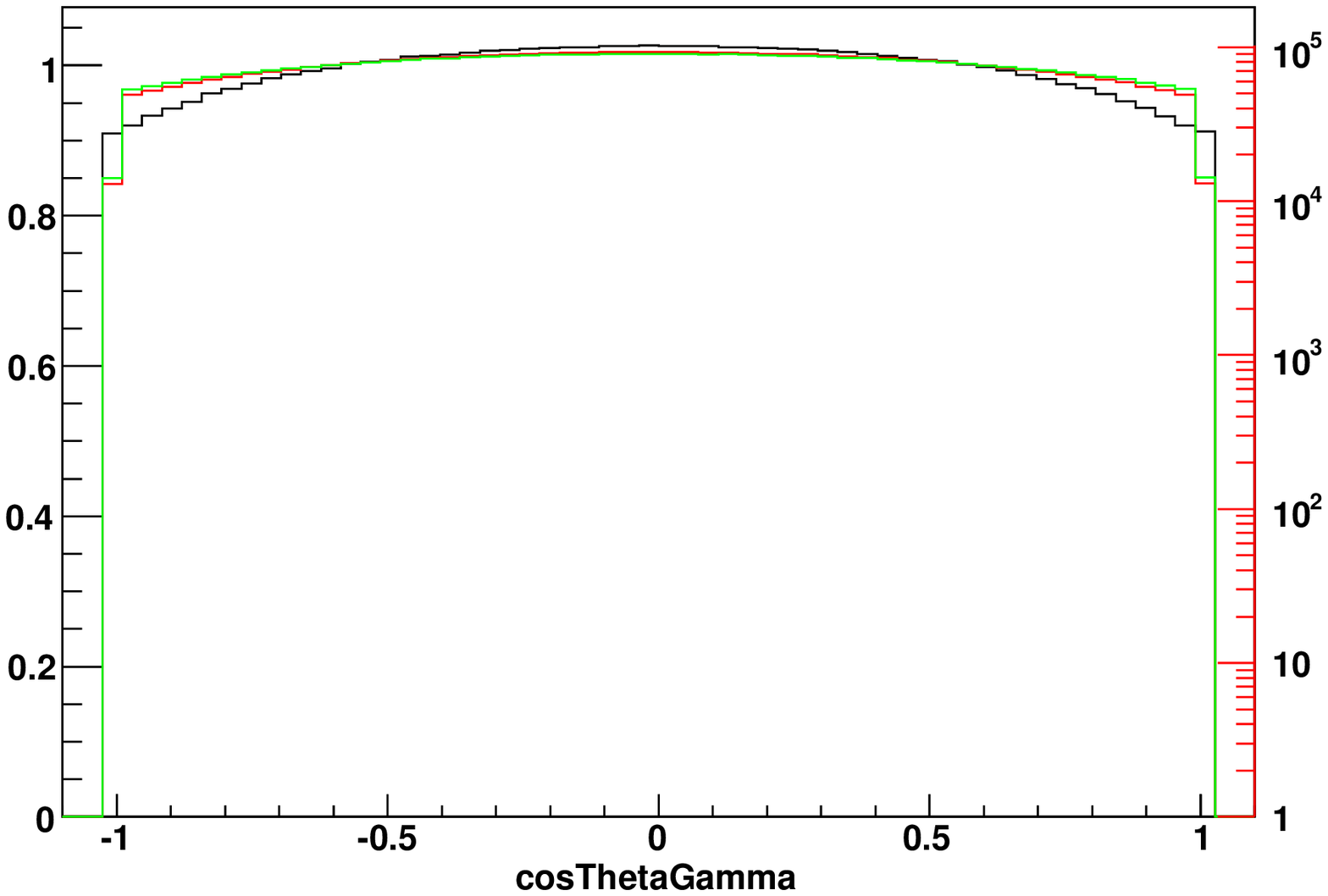}}
\subfigure[ {\it cosine of the $\pi^{+}$ angle with respect to the same charge beam.
}
 ]{\includegraphics[%
 width=0.49\columnwidth,
  keepaspectratio]{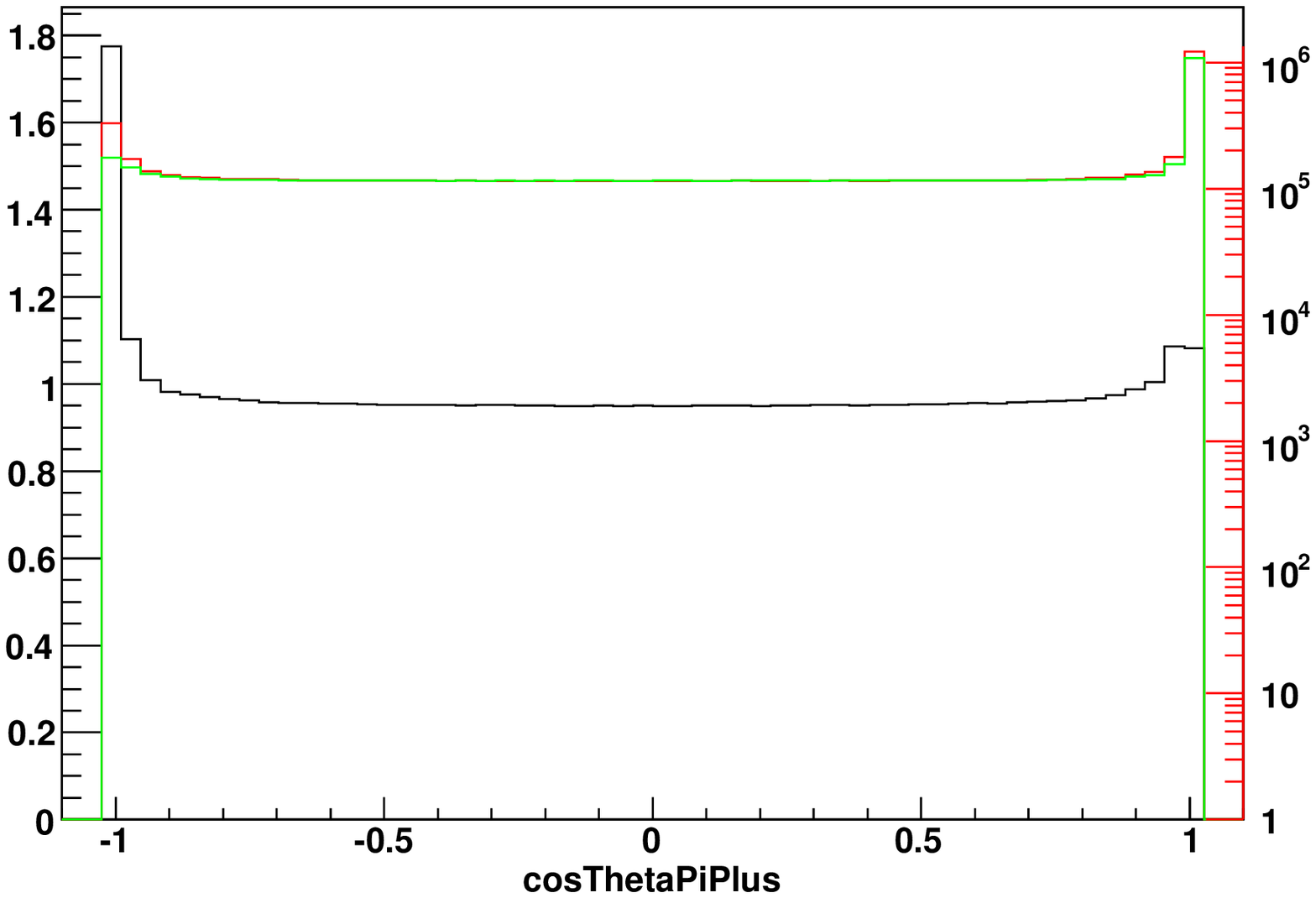}}
\end{center}
\end{figure*}

\begin{figure*}[!h]
\psfrag{cosThetaGamma}{\scriptsize $\cos\theta_{\gamma}$}
\psfrag{cosThetaPiPlus}{\scriptsize $\cos\theta_{\pi+}$}
\caption{\label{fig:oneY} \it \small Angular distributions  for
$e^+e^- \to \pi^+\pi^- (\gamma)$ at 2 GeV center of mass energy. Results 
from  {\tt PHOTOS} with matrix element taken from \cite{Nanava:2006vv} are given in red colour. 
 $ \sum_{\lambda,\epsilon} |M_{I'}^\prime|^2$ 
is  used for results with green line. Logarithmic scale is used, but
for the ratio (black line) linear scale is used instead.
 Fraction of presented events (i.e. with photons above
50 MeV) is respectively 4.2279 $\pm$ 0.0021 \% and 4.2271 $\pm$ 0.0021\% 
of the total samples for the two programs.  
}

\begin{center}
\subfigure[cosine of the photon angle with respect to beam  is shown. 
 ]{\includegraphics[%
  width=0.49\columnwidth,
  keepaspectratio]{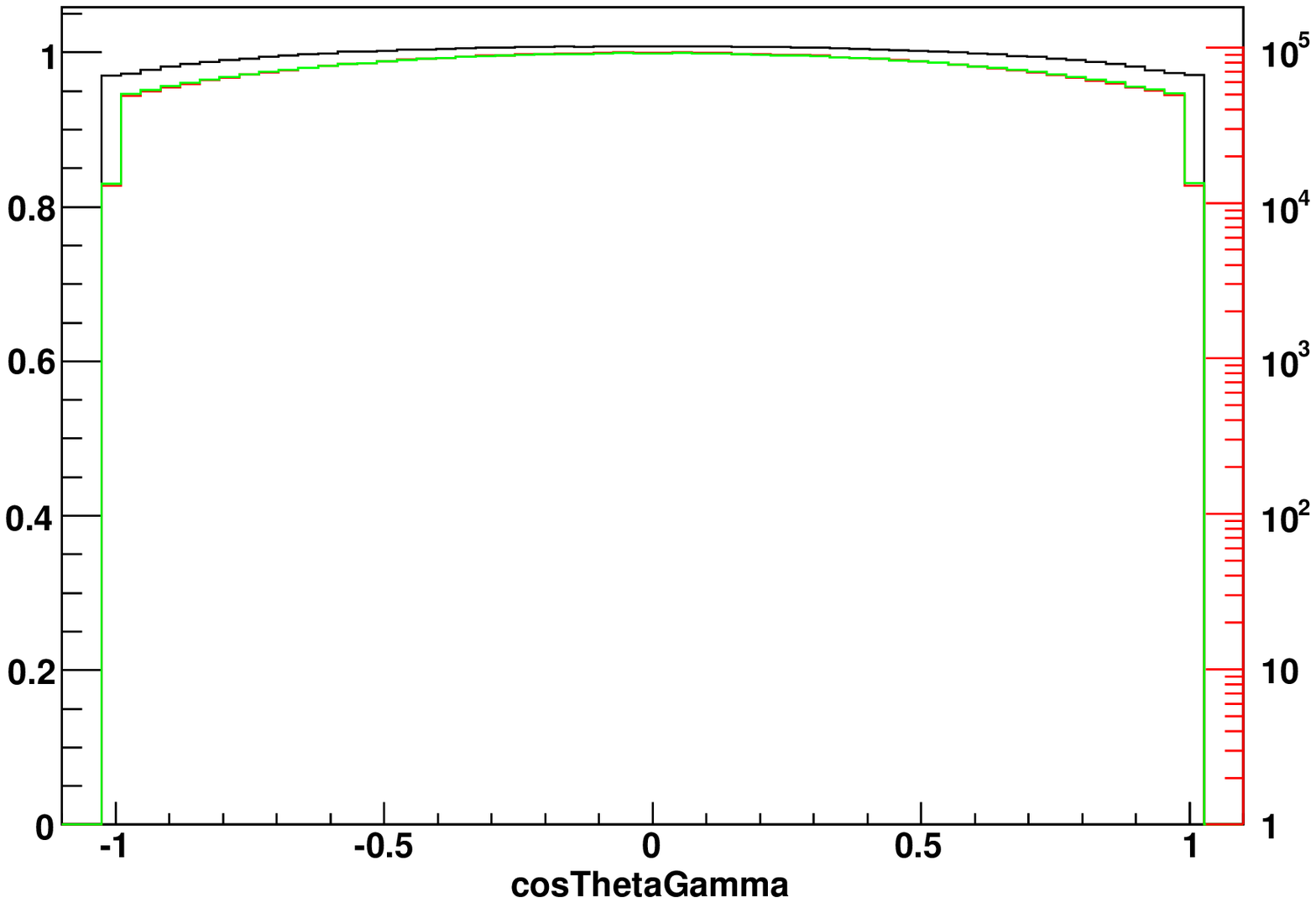}}
\subfigure[ {\it cosine of the $\pi^{+}$ angle with respect to the same charge beam is shown.
}
 ]{\includegraphics[%
 width=0.49\columnwidth,
  keepaspectratio]{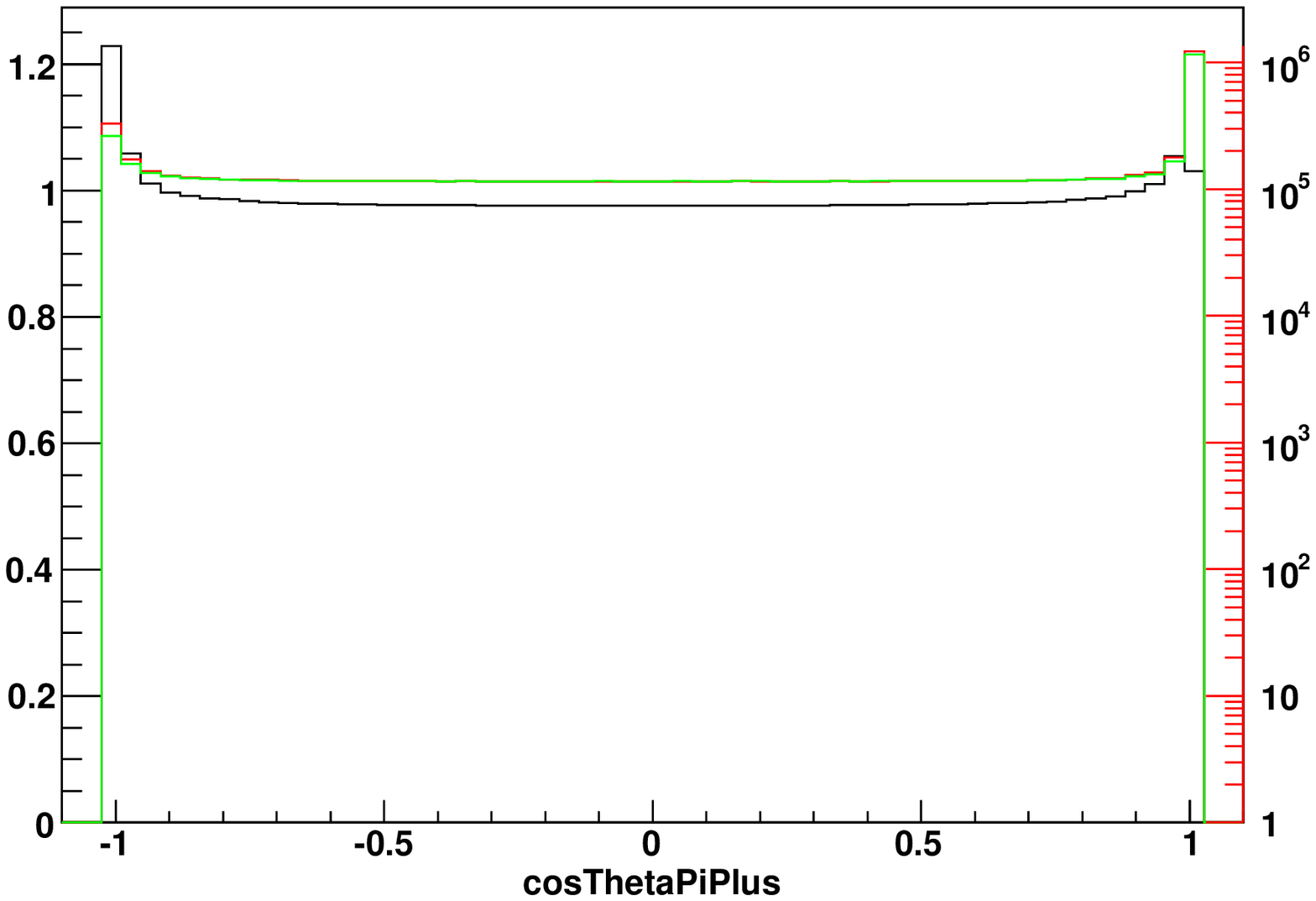}}
\end{center}
\end{figure*}

\section{Summary}

In this paper we have studied matrix elements for the
  $\gamma^* \to \pi^+\pi^- \gamma$ and $W \to l \nu \gamma$ processes. 
We have observed that  the expressions can be separated into gauge invariant parts.
In both  cases,
the part consisting of  eikonal factor multiplying the Born level spin
amplitude separates. This part contributes to the
infrared singularity. 
In the case of  $\gamma^* \to \pi^+\pi^- \gamma$  the remaining part does not
contribute to collinear singularity
but  for $W \to l \nu$ it can be separated further:
into the part proportional to the $W$ charge which does not lead  
to any logarithmic contribution after integration, and the part proportional to
the  lepton charge,
which do contribute to collinear singularity and is identical 
to an analogous part for the $Z \to l^+ l^-$ process 
(see eg. second or third part of formula 5 in \cite{vanHameren:2008dy}). This is exactly the 
factorization property
needed for the iterative solution used in {\tt PHOTOS} 
to be valid for multiphoton emissions and processes discussed in present 
paper.

For  $\gamma^* \to \pi^+\pi^- \gamma$, 
the factor,  identified as Born level amplitude  is not unique.
As expected, ambiguity is proportional to photon momentum, and  disappears in
the soft limit. As a consequence of  the ambiguity
options were discussed.
Their difference is of no practical consequences for the present work, 
but should be kept in mind if the obtained spin amplitude parts
 would be used as building 
bricks for amplitudes of more elaborated processes.

We have identified  dominant parts of spin amplitudes and used them for
tests and for 
 kernels of  {\tt PHOTOS} Monte Carlo. Exact matrix element was used for that
purpose. 
We have found, that the whole matrix element for the process  $W \to l \nu \gamma$ 
can be incorporated into the photon emission kernel. However 
for  $\gamma^* \to \pi^+\pi^- \gamma$ it is not technically straightforward, because
of   large weight events. The responsible term was identified
and it may be interesting to point out that it is
similar to the one obtained in a different calculation (Ref. \cite{vanHameren:2008dy}
equation (68)).
There, such terms  were interpreted as  contributing to running of QCD coupling constant.
In present calculation the term  involves complete kinematics
of  $\gamma^* \to \pi^+\pi^- \gamma$ process.

Any further investigation of the analogy, would require amplitudes of higher 
orders.
 Such  effort can not be justified for scalar QED. 
From low energy point of view,   terms like (\ref{HII}) or (\ref{HIIp})
should be understood as genuine scalar 
QED process, and not part of real photon bremsstrahlung. 
Scalar QED is not supposed 
to be valid in the regions of phase space where these terms contribute 
significantly. 
It is of phenomenological interest to check this limit of 
scalar QED predictions by direct comparisons with  data. 
Thanks to present work, higher order genuine bremsstrahlung effects for
 $\gamma^* \to \pi^+\pi^- \gamma$
can be simulated with the help of {\tt PHOTOS} and one 
can concentrate on confronting  the data with these 
non-bremsstrahlung parts (\ref{HII}) or (\ref{HIIp}).

We have neither discussed here the interference with the photons originating 
from incoming beams, nor the  interference between two consecutive
emissions from the same charged line. The first effect,
requires simultaneous treatment of initial-state  and  final-state 
bremsstrahlung.
This is out of scope of work on {\tt PHOTOS} alone, but
 spin amplitudes are already prepared. For discussion of interference 
of two emissions  from the same charged line (and resulting uncertainties) 
second order matrix element is needed.
Fortunately the structure of spin amplitudes for $\gamma^* \to \pi^+\pi^- \gamma$ and $W\to l \nu$
matches that of 
$Z \to l^+ l^- \gamma$ \cite{Was:2004ig}. At present, we can only expect that 
these results on  $Z \to l^+ l^- \gamma$ in combination with  
our algorithm for matching consecutive emissions,
hold for our $\gamma^* \to \pi^+\pi^- $ processes. Results of Ref. 
\cite{Isidori:2007zt} point that this expectation is well founded.

We have not discussed
virtual corrections. We assume, following Kinoshita-Lee-Nauenberg theorem 
\cite{Kinoshita:1962ur,Lee:1964is}, that the dominant part can be included
in a factor multiplying Born amplitude and the correction to the total rate
is free of any large logarithm.
We leave this point for  future work.

Finally, this paper provides numerical tests of {\tt PHOTOS} Monte Carlo,
in particular
construction for  decays where Born level cross section has a zero.
This is of practical interest for  users of the program and also a necessary step 
before any attempt of an  extension to QCD.


\vskip 2 cm

\appendix
\section{Matrix element for W decay}
 The matrix element of the process  $W^{-}(Q,\lambda) \rightarrow l(p_{l},\lambda_{l})\,\bar{\nu}(p_{\nu},\lambda_{\nu})\, \gamma(k,\sigma)$ has the form 
\begin{eqnarray}
  \textit{M}^{\,\sigma}_{\lambda,\lambda_{\nu},\lambda_{l}}(k,Q,p_{\nu},p_{l}) &=& 
  \left[\frac{Q_{l}}{2\,k\cdot p_{l}}b_{\sigma}(k,p_{l}) -\frac{Q_{W}}{2\,k\cdot Q}\left(b_{\sigma}(k,p_{l})+b_{\sigma}(k,p_{\nu})\right) \right]{\textit{\large B}^{\lambda}_{\lambda_{l},\lambda_{\nu}}(p_{l},Q,p_{\nu})} \nonumber \\
 & & +\frac{Q_{l}}{2\,k\cdot p_{l}}\sum_{\rho=\pm}U^{\sigma}_{\lambda_{l},\rho}(p_{l},m_{l},k,0,k,0) {\textit{\large B}^{\lambda}_{\rho,-\lambda_{\nu}}(k,Q,p_{\nu})} \nonumber\\
 & &-\frac{Q_{W}}{2\,k\cdot Q}\sum_{\rho=\pm}\left({\textit{\large B}^{\lambda}_{\lambda_{l},-\rho}(p_{l},Q,k)} {U^{\sigma}_{-\rho,-\lambda_{\nu}}(k,0,k,0,p_{\nu},0)} \right. \label{AmpW} \\ \nonumber
 & & \left.\phantom{xxxxxxxxxx} + U^{\sigma}_{\lambda_{l},\rho}(p_{l},m_{l},k,0,k,0) {\textit{\large B}^{\lambda}_{\rho,-\lambda_{\nu}}(k,Q,p_{\nu})}\right), \nonumber
\end{eqnarray}
where we use the following notation\,:
\begin{eqnarray}
 \textit{\large B}^{\lambda}_{\lambda_{1},\lambda_{2}}(p_{1},Q,p_{2})  
                     &\equiv& \frac{g}{2\sqrt{2}}\bar{u}(p_{1},\lambda_{1})\,\widehat{\epsilon}^{\lambda}_{W}(Q)(1+\gamma_{5})\,v(p_{2},\lambda_{2})\,,                                                    \nonumber \\
 U^{\sigma}_{\lambda_{1},\lambda_{2}}(p_{1},m_{1},k,0,p_{2},m_{2})
                     &\equiv& \bar{u}(p_{1},\lambda_{1})\,\widehat{\epsilon}^{\sigma}_{\gamma}(k)\,u(p_{2},\lambda_{2})\,,
                                                                                          \\
 \delta_{\lambda_{1}\lambda_{2}}b_{\sigma}(k,p) 
                     &\equiv& U^{\sigma}_{\lambda_{1},\lambda_{2}}(p,m,k,0,p,m) \nonumber \,,
\end{eqnarray}
 $Q_{l}$ and $Q_{W}$ are respectively the electric charges of the fermion $l$ and the $W$ boson, 
in units of the positron charge, ${\epsilon}^{\sigma}_{\gamma}(k)$ and ${\epsilon}^{\lambda}_{W}(Q)$
denote respectively the polarization vectors of the photon and the $W$ boson. An expression of the function $U^{\sigma}_{\lambda_{1},\lambda_{2}}$
in terms of the massless spinors and other notations can be found in~\cite{Jadach:1998jb}.
It is easy to check that the three components  of the sum contributing to (\ref{AmpW}) are individually
gauge invariant.  Note, that the first component coincides with the 
amplitude in the eikonal approximation.

\clearpage

\addcontentsline{toc}{section}{\refname}
\bibliographystyle{utphys_spires}

\bibliography{bibliography}

\end{document}